\documentclass[twocolumn,secnumarabic,amssymb, nobibnotes, aps, prb,superscriptaddress]{revtex4-1}

\usepackage{graphicx}
\usepackage{dcolumn}
\usepackage{bm}
\usepackage{hyperref}
\usepackage{natbib}
\usepackage{amsmath}
\usepackage{amsfonts}

\begin{document}
	
	\preprint{APS/123-QED????????????}
	
	
	\title{On topological transitions in metals} 
	
	\author{Xuzhe Ying}
	\affiliation{School of Physics and Astronomy, University of Minnesota, Minneapolis, MN 55455, USA}
	
	\author{Alex Kamenev}
	\affiliation{School of Physics and Astronomy, University of Minnesota, Minneapolis, MN 55455, USA}
		\affiliation{William I. Fine Theoretical Physics Institute,  University of Minnesota,
Minneapolis, MN 55455, USA}


	
	
	\begin{abstract}
	We investigate if a sharp topological transition in a metal with a large Fermi surface may be detected 
	in transport measurements. In particular, we address if a skew scattering and a side jump on elastic disorder in the bulk  of such a metal masks signatures of the topological transition. We conclude that certain transport coefficients   	exhibit discontinuous changes across the transition.  These discontinuities are not smeared or dwarfed by the bulk metallic transport in a broad range of  parameters.    
	\begin{description}
			\item[PACS numbers]
		\end{description}
	\end{abstract}
	
	\maketitle

\section{Introduction}
\label{sec:intro}

Discovery of topological insulators and semimetals \cite{bernevig2013topological,shen2012topological,RevModPhys.82.3045,RevModPhys.83.1057,RevModPhys.88.035005} emphasized a fundamental fact that states of matter may be distinguished by topological indexes. The existence of topological  indexes relies  on symmetries of the system, rather than a specific Hamiltonian \cite{RevModPhys.88.035005,ryu2010topological,wigner1958distribution,dyson1962threefold,PhysRevB.78.195125,PhysRevX.7.041069}. States with different indexes are separated by topological phase transitions, 
which are often associated with sharp quantized changes of transport coefficients (Hall conductance in the 
integer quantum Hall effect\cite{prange1987quantum} being the oldest example).  Topological transitions are often associated with gapped phases, such as insulators or superconductors. Recent studies of Weyl semimetals \cite{hosur2013recent,RevModPhys.90.015001,RevModPhys.90.015001,PhysRevB.83.205101,xu2015discovery,PhysRevX.5.031013,PhysRevLett.112.136402,PhysRevLett.114.110401} have extended this notion  to gapless states, if the chemical potential is tuned to a nodal Weyl (or Dirac) point \cite{PhysRevB.92.045128,PhysRevLett.114.136801}. For example in Weyl semimetals with mirror symmetry the Hall conductance exhibits a discontinuous quantized change \cite{PhysRevLett.120.016603,PhysRevB.97.165104}. 

In the present work we investigate if sharp topological transitions may be detected in genuine metals with large Fermi surfaces and a finite density of bulk delocalized states. One may reason that the edge states, if coexist with the bulk extended states, provide a negligible contribution to transport coefficients in the thermodynamic limit. Moreover, if elastic scattering  is present it mixes between the edge and the bulk, further degrading any topological signatures.  Here we show that these arguments are too simplistic. We conclude that sharp, but  {\em non-quantized}, discontinuities in transport coefficients persist into the genuine metallic state. Their magnitude is finite and may be comparable with (or even larger than)  bulk contributions in the thermodynamic limit. These properties has to 
be protected by a certain symmetry. We thus refer to them as symmetry protected topological (SPT) {\em metals}. If disorder does not break the corresponding symmetry, the transport anomalies are robust to its presence. 

\begin{figure}[tb]
		\centering
		\includegraphics[width=\linewidth]{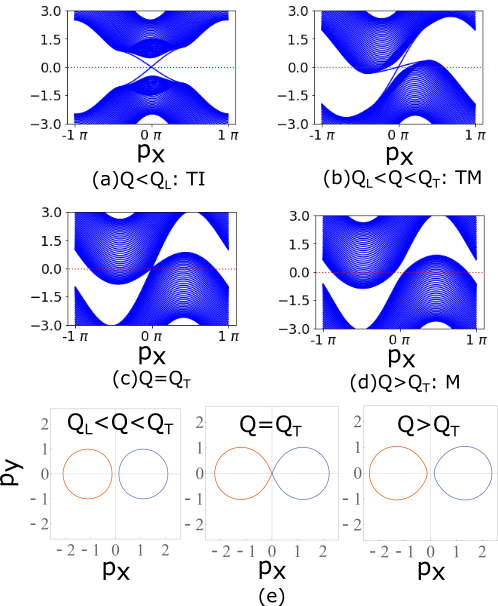}
		\caption{(a)-(d) Band structure of the model from Ref.~[\onlinecite{PhysRevLett.121.086810}]  at various choices of parameter $Q$ (see Sec.~\ref{sec:model} for details of the model). The corresponding phases are: (a) topological insulator/superconductor (TI); (b) topological metal (TM); (c) topological phase transition; (d) normal metal (M). (e) Fermi surfaces, corresponding to cases (b-d) above. Blue lines in panels (a-d) represent multiple sub-bands of transversal quantization in quasi-1D geometry.}
		\label{fig:BandStructures}
	\end{figure}

To illustrate these points we consider a specific model of SPT metal with the particle-hole symmetry, introduced in Ref.~[\onlinecite{PhysRevLett.121.086810}]. The model is based on 2D $p+ip$ superconductor, which is deformed by an applied flux or a super current. As the super current (hereafter denoted as $Q$) is  increased the band structure undergoes two transitions as shown in Fig.~\ref{fig:BandStructures}(a-d). First at $Q=Q_L$, there is a Lifshitz transition from a topological 
superconductor to the topological metal state. The latter is characterized by two metallic  bands with the Fermi surfaces shown in Fig.~\ref{fig:BandStructures}(e). Notice also the two edge states, propagating in the {\em same} direction, coexisting with the metallic bands, Fig.~\ref{fig:BandStructures}(b). At $Q=Q_T>Q_L$, there is the topological transition between the topological and the ordinary metal states. In the clean model the anomalous thermal Hall conductance exhibits a non-quantized discontinuity\cite{PhysRevLett.121.086810} at  $Q=Q_T$.

\begin{figure}[]
		\centering
		\includegraphics[width=\linewidth]{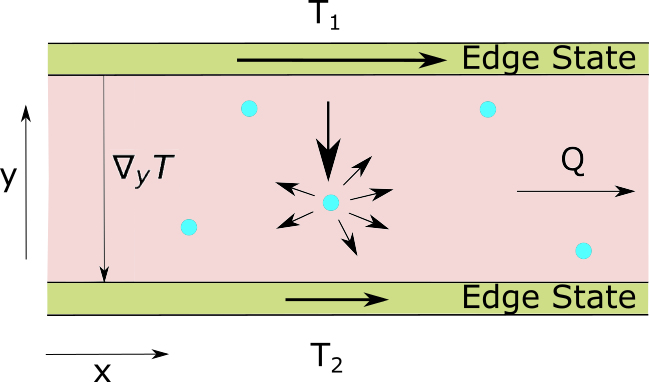}
		\caption{Schematic system setup.  The edge states are in a local equilibrium with the heat reservoirs at  temperatures $T_1$ and $T_2$.  The temperature gradient is established in the bulk of the system, leading to the quasiparticles drift. The anisotropic (skew) scattering on impurities (blue dots) leads to a heat current in $x$-direction, indicated by the arrows around the blue dot. Notice that in the topological metal state both edges propagate in the same direction, cf. Fig.~\ref{fig:BandStructures}(b).}
		\label{fig:setup}
	\end{figure}

In the present paper we investigate thermal transport in this model in the presence of a bulk elastic disorder, which preserves the particle-hole symmetry.  The schematic setup is indicated in Fig.~\ref{fig:setup}. It assumes an applied thermal gradient, $\nabla_y T=(T_2-T_1)/L_y$ in the $y$-direction. We then evaluate linear response thermal current both in 
$y$ and $x$-directions, which determine longitudinal, $\kappa_{yy}$, and Hall, $\kappa_{xy}$, thermal conductances.  The latter is, in general, non-zero, because the time-reversal invariance is broken in two ways: (i) the choice of chirality of 
$p+ip$ order parameter and (ii) the applied super current $Q$.  As illustrated in Fig.~\ref{fig:setup}, $\kappa_{xy}$ acquire contributions from both edge states and bulk elastic scattering.

The main focus of this paper is whether the discontinuity in $\kappa_{xy}$, found in the clean model\cite{PhysRevLett.121.086810}, Fig.~\ref{fig:BP}, is observable in the presence of the bulk disorder.  We found that this is indeed the case.
The skew scattering mechanism results in a continuous contribution to $\kappa_{xy}$, while the side jump mechanism enhances the discontinuity of the intrinsic contribution. Moreover, in the wide range of parameters the continuous skew scattering contribution is comparable to the discontinuity size. The discontinuity is not smeared by the disorder in the thermodynamic limit.  (The discontinuity in $\kappa_{xy}$ should be understood in a scaling limit when $T_1$ and $T_2$ go to zero, while  a finite temperature leads to a shop by continuous drop in $\kappa_{xy}$.)

	The structure of the paper is as follows: Section \ref{sec:model} introduces the model of SPT metal and discusses the intrinsic contribution to the thermal Hall conductance. Section \ref{sec:kinetic} establishes the appropriate kinetic Boltzmann equation for impurity scattering of quasiparticles occupying particle-like and hole-like conduction bands. Section\ref{sec:results} presents our results and discussion of the topological signatures in the transport coefficients. The rest of the paper provides the necessary details to support the main results: Appendix \ref{sec:scattering}-\ref{app:rates} is devoted to a systematic evaluation of the impurity scattering rates; Appendix \ref{sec:SideJump}-\ref{app:SJCal} is devoted to the side jump displacement and side jump velocities as well as its correction to the thermal Hall conductance.

\section{Model of SPT metal}
\label{sec:model}	
	
We consider a model of SPT metal\cite{PhysRevLett.121.086810}, based on $2$-D $p+ip$ superconductor deformed by an applied supercurrent. It is described by the Bogoliubov-de Gennes (BdG) Hamiltonian:
	\begin{equation}
	H_{\mathrm{BdG}}(\boldsymbol{p}) = \begin{pmatrix}
	\xi_{p_x+Q/2,p_y}  & \Delta(\boldsymbol{p})\\
	\Delta^*(\boldsymbol{p})  	 & -\xi_{p_x-Q/2,p_y}
	\end{pmatrix}, 
	\label{TBH}
	\end{equation}
where 
\begin{equation}
\xi_{p_x,p_y}=\frac{p_x^2+p_y^2}{2m}-\mu,
\end{equation}  
is electron dispersion close to the bottom of a band and $\mu$ is the chemical potential.  The dispersion relation is deformed by the applied flux (or the supercurrent) assumed to be in the $x$-direction and denoted as $Q$. Notice that it enters the BdG Hamiltonian as a canonical substitution $p_x \to p_x+{Q\over 2} \sigma_z$, where $\sigma_z$ is the Pauli matrix in the particle-hole space. 

The off-diagonal part of Eq.~(\ref{TBH}) is the $p$-wave superconducting gap function. In the continuous limit, adopted here, it is given by 
	\begin{equation}
	\begin{split}
	\Delta(\boldsymbol{p})=-2\Delta(p_y+ip_x),
	\end{split}
	\label{CH}
	\end{equation}
where $\Delta$ is $p$-wave paring amplitude (notice that $\Delta$ has a dimensionality of velocity). Throughout the paper we assume that the superconductivity is proximity induced and thus do not try to establish a value of $\Delta$ through a self-consistency relation. We also do not discuss a possible  suppression of $\Delta$ by disorder.

\begin{figure}[tb]
		\centering
		\includegraphics[width=\linewidth]{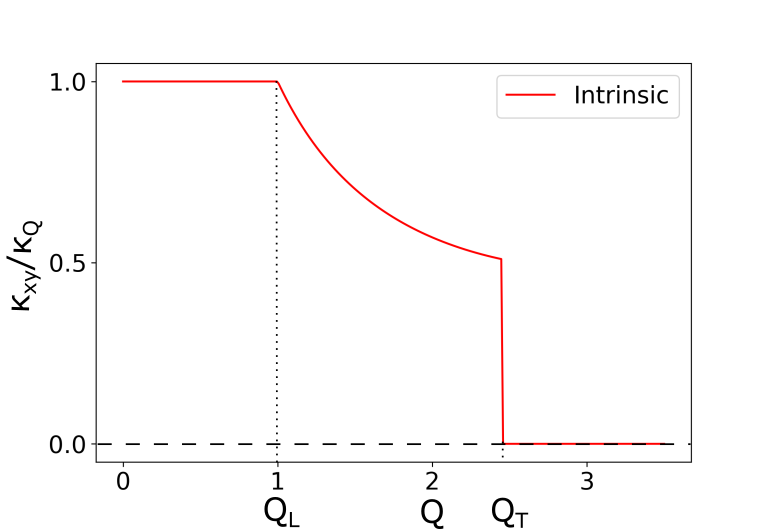}
		\caption{The intrinsic contribution to the thermal Hall conductivity in units of $\kappa_Q=(\pi k_B^2/12\hbar)T$ as a function of the suppercurrent  $Q$ (measured in units of $4m\Delta$). The calculation is done for $v_F/\Delta=4.9$.}
		\label{fig:BP}
	\end{figure}
	

It is instructive to rewrite the Hamiltonian (\ref{TBH}) in the matrix notations:
	\begin{equation}
	\begin{split}
	&H_{\mathrm{BdG}}(\boldsymbol{p})=d_0(\boldsymbol{p})+\boldsymbol{d}(\boldsymbol{p})\cdot\boldsymbol{\sigma};\\
	&d_0(\boldsymbol{p}) = Qp_x/2m;\\
	&d_x(\boldsymbol{p})=-2\Delta p_y\ \ \ d_y(\boldsymbol{p})=-2\Delta p_x;\\
	&d_z(\boldsymbol{p})=\boldsymbol{p}^2/2m+Q^2/8m-\mu,
	\end{split}
	\label{Eq:Ham}
	\end{equation}
	where $\boldsymbol{\sigma}$ is a vector of the Pauli matrices in the particle-hole space. This Hamiltonian has two quasiparticle bands with the energies, Fig.~\ref{fig:BandStructures}:
	\begin{equation}
	\epsilon^{(s)}_{\boldsymbol{p}}=d_0(\boldsymbol{p})+s |\boldsymbol{d}(\boldsymbol{p})|;\ \ \ \ \ (s=\pm),
	\label{Eq:QPE}
	\end{equation}
	where $|\boldsymbol{d}(\boldsymbol{p})|=\sqrt{d^2_x(\boldsymbol{p})+d^2_y(\boldsymbol{p})+d^2_z(\boldsymbol{p})}$. Hereafter $s$($s^{\prime}$) is the label for the quasiparticle bands and takes the values of $\pm$. The corresponding quasiparticle wavefunctions (Nambu spinors) are independent on $d_0(\boldsymbol{p})$ and may be expressed through $\boldsymbol{d}(\boldsymbol{p})$ as
	\begin{equation}
	\begin{split}
	&\psi^{(+)}_{\boldsymbol{p}}=\mathcal{N}_{\boldsymbol{p}}\begin{pmatrix}
	|\boldsymbol{d}(\boldsymbol{p})|+d_z(\boldsymbol{p})\\
	d_x(\boldsymbol{p})-id_y(\boldsymbol{p})
	\end{pmatrix}\\
	&\psi^{(-)}_{\boldsymbol{p}}=\mathcal{N}_{\boldsymbol{p}}\begin{pmatrix}
	-d_x(\boldsymbol{p})-id_y(\boldsymbol{p})\\
	|\boldsymbol{d}(\boldsymbol{p})|+d_z(\boldsymbol{p})
	\end{pmatrix}
	\end{split}
	\label{Eq:QPWF}
	\end{equation}
	where $\mathcal{N}_{\boldsymbol{p}}= \big[2|\boldsymbol{d}(\boldsymbol{p})|(|\boldsymbol{d}(\boldsymbol{p})|+d_z(\boldsymbol{p}))\big]^{-1/2}$ is a normalization factor. 
	
	The two critical points, see Fig.~\ref{fig:BandStructures},  may be determined from the dispersion relation (\ref{Eq:QPE}). The two Fermi surfaces form at the Lifshitz transition $Q_L=4m\Delta$, when zero energy solution $\epsilon^{(s)}_{\boldsymbol{p}}=0$ first appear. The topological phase transition takes place at $Q_T=2k_F=2\sqrt{2m\mu}$, when the two bands touch at $\boldsymbol{p}=0$, because $|\boldsymbol{d}(0)|=0$.

	The Hamiltonian of Eq.~(\ref{TBH}) possesses  the particle-hole symmetry, which is expressed as  

	\begin{equation}
					\label{eq:particle-hole}
	P^{-1}H_{\mathrm{BdG}}(\boldsymbol{p})P=-H_{\mathrm{BdG}}(-\boldsymbol{p}),
	\end{equation}
	where the particle-hole symmetry operator is defined as $P=\sigma_xK$, where $K$ is complex conjugation operator.
	
We study thermal Hall conductance of the system described by the Hamiltonian (\ref{TBH}) in the geometry indicated in Fig.~\ref{fig:setup}. The thermal gradient $\nabla_yT$ is applied in the $y$ direction, while the supercurrent $Q$ flows in the $x$-direction. In the linear response the heat current is given by $j^{\mathrm{heat}}_x=\kappa_{xy}\nabla_yT$, where $\kappa_{xy}$ is the thermal Hall conductance. It has two contributions: the intrinsic one due to the edge states and the impurity one due to the skew-scattering.

According to the Kubo-St\u reda formula \cite{read2000paired}, the  
intrinsic (anomalous) thermal Hall conductance (in unit of $\kappa_Q= \frac{\pi k_B^2}{12\hbar}T$) is given by the integrated Berry curvature:	
	\begin{equation}
							\label{eq:intrinsic}
	\kappa_{xy}^{\mathrm{int}}=\kappa_Q\sum_{s=\pm}\int_{BZ}\frac{d^2p}{(2\pi)^2}\, n^0(\epsilon^{(s)}_{\boldsymbol{p}})\, \Omega_z^{(s)}(\boldsymbol{p}),
	\end{equation}
where the momentum integration runs over the Brillouin zone (BZ) of a proper lattice model\cite{PhysRevLett.121.086810}.	 
In the superconducting phase $Q<Q_L$ the intrinsic conductance is quantized $\kappa_{xy}^{\mathrm{int}}=\kappa_Q$, Fig.~\ref{fig:BP}. In the topological metal phase, $Q_L<Q<Q_T$, the quantization is broken by the fact that the Fermi surfaces need to be exempt from the BZ summation (at small temperature). At the topological phase transition $Q\geq Q_T$ the intrinsic Hall conductivity discontinuously jumps to zero. The discontinuity in the Hall conductance is a consequence of (and is protected by) the particle-hole symmetry, Eq.~(\ref{eq:particle-hole}). These statements have a natural geometric interpretation in terms of the flux of the Berry monopole, as explained in Ref.~[\onlinecite{PhysRevLett.121.086810}].  

We turn now to the contribution to the thermal Hall conductance due to the scattering of bulk quasiparticles on an elastic disorder.

	\section{Kinetic equation treatment of elastic scattering}
	\label{sec:kinetic}

	\subsection{Review on Anomalous Hall Effect}
	
	Our treatment of the thermal Hall conductance of SPT metals shares some resemblance to the discussion of the anomalous Hall effect (AHE) in ferromagnetic metals\cite{RevModPhys.82.1539,nagaosa2006anomalous}. 	
	It was pioneered by Karplus and Luttinger\cite{karplus1954hall} who showed that a specific structure of Bloch wave functions  gives rise to an anomalous Hall conductivity, which is known as the {\em intrinsic} contribution. Soon after, Smit\cite{smit1955spontaneous,smit1958spontaneous} pointed out that impurities should play an important role in AHE through anisotropic scattering, which was called a {\em skew} scattering. Subsequently systematic diagrammatic calculations\cite{PhysRevB.75.045315,PhysRevB.78.060501,PhysRevB.80.104508,PhysRevLett.118.027001,ado2015anomalous} were developed to include both intrinsic and impurity scattering contribution to the AHE. It was shown that certain diagrams, e.g. `Mercedes star' diagram\cite{PhysRevB.75.045315,PhysRevB.78.060501,PhysRevB.80.104508,PhysRevLett.118.027001} or `X' and `$\Psi$' diagrams\cite{ado2015anomalous,PhysRevLett.118.027001}, containing higher order terms in Born series must be included to see the effect of skew scattering. At the meantime, it was noticed that the ladder diagrams contributions are on the order of $\mathcal{O}(1)$ independent of impurity strength\cite{RevModPhys.82.1539}. This contribution was named as {\em side jump}, indicating that the shift of center of wave packets at impurity scattering events \cite{PhysRevB.73.075318,RevModPhys.82.1539}. 
	
	Alternatively, one can take a kinetic approach based on semiclassical Boltzmann transport equation\cite{PhysRevB.75.045315,sinitsyn2007semiclassical,konig2018gyrotropic}, which is equivalent to diagrammatic resummation and typically is more intuitive. 
	In such kinetic approach, the nature of the Bloch waves is taken into account by the so-called anomalous velocity in the semiclassical equation of motion for a Bloch electron\cite{PhysRevB.59.14915}:
	\begin{equation}
	\begin{split}
	&\dot{\boldsymbol{r}}=\boldsymbol{\nabla}_{\boldsymbol{p}}\epsilon_{\boldsymbol{p}} + \dot{\boldsymbol{p}}\times\boldsymbol{\Omega}(\boldsymbol{p})\\
	&\dot{\boldsymbol{p}}=-\boldsymbol{\nabla}_{\boldsymbol{r}}U(\boldsymbol{r})
	\end{split}
	\end{equation}
	Here, the Bloch electron is subject to an external  force $\boldsymbol{F}=-\boldsymbol{\nabla}_{\boldsymbol{r}}U(\boldsymbol{r})$ and $\boldsymbol{v}^{\mathrm{G}}_{\boldsymbol{p}}=\boldsymbol{\nabla}_{\boldsymbol{p}}\epsilon_{\boldsymbol{p}}$ is the usual group velocity, while $\boldsymbol{v}^{\mathrm{A}}_{\boldsymbol{p}}=\dot{\boldsymbol{p}}\times\boldsymbol{\Omega}(\boldsymbol{p})= \boldsymbol{F}\times\boldsymbol{\Omega}(\boldsymbol{p})$ is the anomalous velocity.  Here $\boldsymbol{\Omega}(\boldsymbol{p})=\boldsymbol{\nabla}_{\boldsymbol{p}}\times\boldsymbol{\mathcal{A}}(\boldsymbol{p})$ is the Berry curvature, given by momentum space curl of Berry connection, $\boldsymbol{\mathcal{A}}(\boldsymbol{p})=\langle u(\boldsymbol{p})|i\boldsymbol{\nabla}_{\boldsymbol{p}}|u(\boldsymbol{p})\rangle$. Here, $|u(\boldsymbol{p})\rangle$ is the periodic part of the Bloch wavefunction.
	
Impurity scattering may be described by the semiclassical Boltzmann equation with a proper collision integral.   It's important that all orders in Born series for impurity scattering are taken into account by the Lippmann-Schwinger equation\cite{PhysRevB.75.045315,sinitsyn2007semiclassical,PhysRevB.92.100506}. In the linear response one can solve the Boltzmann equation to find a stationary distribution function: $
	n_{\boldsymbol{p}}=n^0(\epsilon_{\boldsymbol{p}})+\delta n_{\boldsymbol{p}}$, 
	where $n^0(\epsilon_{\boldsymbol{p}})$ is the equilibrium distribution function and $\delta n_{\boldsymbol{p}}\propto\mathbf{F} $  is the deviation from equilibrium due to the external force. 	Finally the current is given by: 
	\begin{equation}
	\boldsymbol{j}^{\prime}=\int d\boldsymbol{p}(\boldsymbol{v}^{\mathrm{G}}_{\boldsymbol{p}}+\boldsymbol{v}^{\mathrm{A}}_{\boldsymbol{p}})[n^0(\epsilon_{\boldsymbol{p}})+\delta n_{\boldsymbol{p}}] = \boldsymbol{j}^{\mathrm{int}}+\boldsymbol{j}^{\mathrm{skew}},
	\end{equation}
	where in the linear response $\boldsymbol{j}^{\mathrm{int}}=\int d\boldsymbol{p}\ \boldsymbol{v}^{\mathrm{A}}_{\boldsymbol{p}}n^0(\epsilon_{\boldsymbol{p}})$ and $\boldsymbol{j}^{\mathrm{skew}}=\int d\boldsymbol{p}\ \boldsymbol{v}^{\mathrm{G}}_{\boldsymbol{p}}\delta n_{\boldsymbol{p}}$. Notice that since the anomalous velocity is already proportional to the external force, the intrinsic contribution relies only on the equilibrium distribution $n^0$. 
	
In addition, for a Bloch electron, the side jump effects will give additional contributions to the current\cite{PhysRevB.73.075318}:
	\begin{equation}
	\boldsymbol{j}^{\mathrm{Side\ Jump}}=\boldsymbol{j}^{\mathrm{sj}}+\boldsymbol{j}^{\mathrm{adist}}.
	\end{equation}
	This is a result of the center of a wave packet experiencing a displacement upon impurity scattering events. Such a  displacement, called the side jump displacement, $\delta\boldsymbol{r}^{\mathrm{sj}}$, depends on the scattering incoming and outgoing states. Here, we briefly reproduce the side jump treatment, following Ref.~[\onlinecite{PhysRevB.73.075318}].
	
	One effect of the side jump is that the quasiparticle velocity acquires an additional component. Qualitatively, at each scattering events, the center of a wave packets is shifted by the side jump displacement $\delta\boldsymbol{r}^{\mathrm{sj}}$. Impurity scattering occurs at a rate of the inverse mean free time $\tau^{-1}$. As a result, the quasiparticles acquire the side jump velocity $
	\boldsymbol{v}^{\mathrm{sj}}_{\boldsymbol{p}}\sim \delta\boldsymbol{r}^{\mathrm{sj}}\tau^{-1}$, leading to a contribution to the current of the form: 
	\begin{equation}
	\boldsymbol{j}^{\mathrm{sj}}=\int d\boldsymbol{p}\ \boldsymbol{v}^{\mathrm{sj}}_{\boldsymbol{p}}\ \delta n_{\boldsymbol{p}}.
	\end{equation}
	
	The second contribution, $\boldsymbol{j}^{\mathrm{adist}}$, originates from the fact that the side jump displacement requires an extra work performed by the external force.  As a result, the quasiparticle energy changes by $
	\delta\epsilon\sim\boldsymbol{F}\cdot\boldsymbol{r}^{\mathrm{sj}}$, leading to an `anomalous' correction, 
	$g^{\mathrm{a}}_{\boldsymbol{p}}$, to a stationary distribution function (see below).  This contributes to the current as:
	\begin{equation}
	\boldsymbol{j}^{\mathrm{adist}}=\int d\boldsymbol{p}\ \boldsymbol{v}^{\mathrm{G}}_{\boldsymbol{p}}\ g^{\mathrm{a}}_{\boldsymbol{p}},
	\end{equation}
	while the total current acquires the form: 
	\begin{equation}
	\boldsymbol{j}=\boldsymbol{j}^{\mathrm{int}}+\boldsymbol{j}^{\mathrm{skew}}+\boldsymbol{j}^{\mathrm{sj}}+\boldsymbol{j}^{\mathrm{adist}}.
	\end{equation}
	
	This scheme refers to the electric current, rather than to the thermal one. To access the latter the Wiedemann-Franz law may be  employed for noninteracting system as shown by Qin, Niu and Shi\cite{PhysRevLett.107.236601} based on the Luttinger's theory of the thermal transport\cite{PhysRev.135.A1505}. Here we employ a kinetic theory of  $p+ip$ superconductors recently developed by Li, Andreev and Spivak\cite{PhysRevB.92.100506}. These authors were interested in a transport of thermally excited quasiparticles in the gapped regime. We adopt their approach to the low-temperature transport in the deformed $p+ip$ model with the two Fermi surfaces.

	\subsection{Kinetic Equation}

\begin{figure}[tb]
		\centering
		\includegraphics[width=\linewidth]{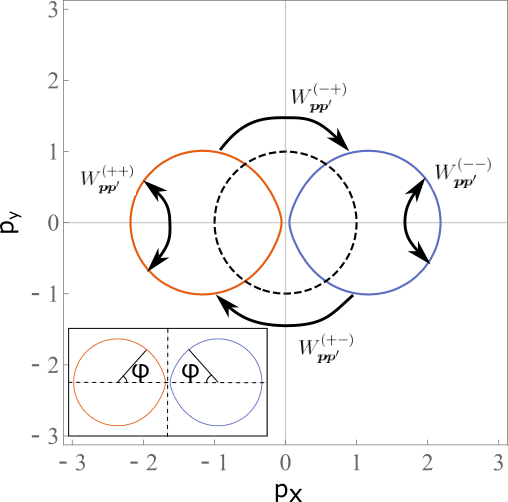}
		\caption{Particle (orange) and hole (blue) Fermi surfaces close to the topological phase transition; various scattering processes with scattering rates $W^{(ss^{\prime})}_{\boldsymbol{p}\boldsymbol{p}^{\prime}}$ are indicated by arrows; dashed circle of radius $4m\Delta$ separates two distinct contributions to $\sum_\mathbf{k} \hat G_0$. Inset: the same Fermi surfaces parameterized by the angle $\varphi$.}
		\label{fig:FSS}
	\end{figure}

    Here we develop a semiclassical Boltzmann equation approach for the quasiparticles with the energies and wavefunctions given by Eqs.~(\ref{Eq:QPE}) and (\ref{Eq:QPWF}) correspondingly. The applicability of this approach is limited to a weak disorder, where the longitudinal conductance is much larger than the conductance quantum. At low temperature this is only possible at $Q>Q_L$.  In this regime the two Fermi surfaces are present and a weak disorder induces elastic scattering between states close to the Fermi surfaces as shown in Fig.~\ref{fig:FSS}.   

	Within each band, labeled as $s=\pm$, the quasiparticle states are assigned a distribution function, $n_{\boldsymbol{p}}^{(s)}$.  As in Fig.~\ref{fig:FSS}, the elastic scattering induces {\em intra} as well as {\em inter} band transitions, which enter the Boltzmann equation through the collision integrals (see eg. Ref.~[\onlinecite{Kamenev2011}]). The coupled Boltzmann equations for the two bands  take the form:
	\begin{equation}
	\frac{\partial n_{\boldsymbol{p}}^{(s)}}{\partial t} + \boldsymbol{v}_{\boldsymbol{p}}^{(s)}\cdot\frac{\partial n_{\boldsymbol{p}}^{(s)}}{\partial \boldsymbol{r}}+\dot{\boldsymbol{p}}\cdot\frac{\partial n_{\boldsymbol{p}}^{(s)}}{\partial \boldsymbol{p}}=I_{\boldsymbol{p}}^{(s)}[n^{(\pm)}]
	\label{Eq:BE}
	\end{equation}
	where $\boldsymbol{v}_{\boldsymbol{p}}^{(s)}=\partial\epsilon^{(s)}_{\boldsymbol{p}}/\partial\boldsymbol{p}$ is quasiparticle velocity. Notice that the anomalous velocity due to Berry curvature\cite{PhysRevB.59.14915} is not included here since it gives rise to the intrinsic contribution of Hall conductivity and does not mix with impurity contribution in the limit of linear response. 
	
 	As indicated in Fig.~\ref{fig:FSS}, the collision integrals involve several scattering processes between quasiparticle states: scattering within each band with scattering rates $W^{(++)}_{\boldsymbol{p}\boldsymbol{p}^{\prime}}$ and $W^{(--)}_{\boldsymbol{p}\boldsymbol{p}^{\prime}}$; scattering between two bands with scattering rates $W^{(+-)}_{\boldsymbol{p}\boldsymbol{p}^{\prime}}$ and $W^{(-+)}_{\boldsymbol{p}\boldsymbol{p}^{\prime}}$. Thus the elastic collision integrals take the form:
	\begin{equation}
	\begin{split}
	I^{(s)}_{\boldsymbol{p}}=&\sum_{s^{\prime}=\pm}\int d\Gamma^{\prime}
	\delta(\epsilon^{(s^{\prime})}_{\boldsymbol{p}^{\prime}}-\epsilon^{(s)}_{\boldsymbol{p}}) \\
	&\times[W^{(ss^{\prime})}_{\boldsymbol{p}\boldsymbol{p}^{\prime}}n^{(s^{\prime})}_{\boldsymbol{p}^{\prime}}(\boldsymbol{r}^{\prime})-W^{(s^{\prime}s)}_{\boldsymbol{p}^{\prime}\boldsymbol{p}}n^{(s)}_{\boldsymbol{p}}(\boldsymbol{r})], 
	\end{split}
	\label{Eq:CI}
	\end{equation}
	where the integral is defined as $d\Gamma^{\prime}=d^2p^{\prime}/(2\pi\hbar)^2$. Here, in the collision integral, the two distribution functions are evaluated at a slightly different location. The difference in position is given by the side jump displacement:
	\begin{equation}
	\boldsymbol{r}^{\prime}-\boldsymbol{r}=\delta\boldsymbol{r}^{\mathrm{sj},(s^{\prime}s)}_{\boldsymbol{p}^{\prime}\boldsymbol{p}}.
	\end{equation}
	
	We now assume that a small temperature gradient is applied in $y$-direction and look for the deviations of the distribution functions from their equilibrium form:
	\begin{equation}
	n^{(s)}_{\boldsymbol{p}}=n^0(\epsilon^{(s)}_{\boldsymbol{p}})+\delta n^{(s)}_{\boldsymbol{p}}+g^{\mathrm{a},(s)}_{\boldsymbol{p}},
	\label{eq:DistFunc}
	\end{equation}
	where $n^0(\epsilon^{(s)}_{\boldsymbol{p}})$ is the equilibrium Fermi distribution. The deviations from equilibrium Fermi distribution can be separated into two part: (i) $\delta n^{(s)}_{\boldsymbol{p}}$ is the solution of the Boltzmann equation with elastic collision by neglecting the side jump effect; (ii) $g^{\mathrm{a},(s)}_{\boldsymbol{p}}$ is called the `anomalous distribution function', compensating the effect of the position change (side jump) over the impurity scattering events. These two parts can be solved independently, due to the linearity of the collision integral, Eq.~(\ref{Eq:CI}) (and the Boltzmann equation, Eq.~(\ref{Eq:BE})). Since we restricted ourselves to the linear response, we only seek for the deviations $\delta n^{(s)}_{\boldsymbol{p}}$ and $g^{\mathrm{a},(s)}_{\boldsymbol{p}}$ proportional to the temperature gradient.

	\subsection{Skew Scattering}
	
	Neglecting for now the side jump effect, the collision integral, Eq.~(\ref{Eq:CI}), reads:
	\begin{equation}
	\begin{split}
	I^{\mathrm{skew},(s)}_{\boldsymbol{p}}=&\sum_{s^{\prime}=\pm}\int d\Gamma^{\prime}
	\delta(\epsilon^{(s^{\prime})}_{\boldsymbol{p}^{\prime}}-\epsilon^{(s)}_{\boldsymbol{p}}) \\
	&\times[W^{(ss^{\prime})}_{\boldsymbol{p}\boldsymbol{p}^{\prime}}\delta n^{(s^{\prime})}_{\boldsymbol{p}^{\prime}}-W^{(s^{\prime}s)}_{\boldsymbol{p}^{\prime}\boldsymbol{p}}\delta n^{(s)}_{\boldsymbol{p}}], 
	\end{split}
	\label{Eq:SKCI}
	\end{equation}

	The effect of skew scattering manifested itself in the asymmetry of the scattering rate. Namely, in the absence of the time-reversal symmetry the matrix elements are not symmetric:
	\begin{equation}
						\label{eq:not-symmetry}
 	W^{(ss')}_{\boldsymbol{p}\boldsymbol{p}^{\prime}} \neq W^{(s's)}_{\boldsymbol{p}^{\prime}\boldsymbol{p}} 
	\end{equation}
	Or more explicitly (as will be derived in Appendix \ref{sec:scattering}-\ref{app:rates}), the scattering rates have the following form\cite{PhysRevB.92.100506}:
	\begin{equation}
		W^{(ss^{\prime})}_{\boldsymbol{p}\boldsymbol{p}^{\prime}}=W^{0,(ss^{\prime})}_{\boldsymbol{p}\boldsymbol{p}^{\prime}}+W^{1,(ss^{\prime})}_{\boldsymbol{p}\boldsymbol{p}^{\prime}}\cos(\theta_{\boldsymbol{p}^{\prime}}-\theta_{\boldsymbol{p}}-2\delta_Q), 
		\label{eq:Wformula}
	\end{equation}
	where the azimuthal angle $\theta_{\boldsymbol{p}}$ ($\theta_{\boldsymbol{p}^{\prime}}$) defines the direction of the momentum $\boldsymbol{p}$ ($\boldsymbol{p}^{\prime}$). And $\delta_Q$ is a small deflection angle, which depends on the metallic density of states $\nu$ and the impurity potential $V_0$ (see Eq.~(\ref{Eq:DA}) in Appendix \ref{sec:scattering}):
	\begin{equation}
	\tan \delta_Q\approx\pi\nu V_0.
	\label{eq:deltaQ}
	\end{equation}
	Its presence is crucial for the asymmetrical property Eq.~(\ref{eq:not-symmetry}), given that the prefactors $W^{0/1,(ss^{\prime})}_{\boldsymbol{p}\boldsymbol{p}^{\prime}}$ are symmetric.
	
	Therefore the collision integral does not conform to the detailed balance. As a result,  it can't be written in the form  
	$W_{\boldsymbol{p}\boldsymbol{p}^{\prime}}n_{\boldsymbol{p}^{\prime}}(1-n_{\boldsymbol{p}}) -    W_{\boldsymbol{p}'\boldsymbol{p}}n_{\boldsymbol{p}}(1-n_{\boldsymbol{p}'})$. Nevertheless the form (\ref{Eq:SKCI}), linear in the distribution functions, is the correct one and it is this form that follows from the Keldysh formalism\cite{0038-5670-27-11-A06,Kamenev2011}\footnote{We are indebted to Anton Andreev for discussions of this issue}.  The collision integral still vanishes at equilibrium, $n^{(s)}_{\boldsymbol{p}} = n^0(\epsilon^{(s)}_\mathbf{p})$, where $n^0(\epsilon)$ is the Fermi function. This is enforced  by the unitarity condition\cite{landau1981course}: 

	\begin{equation}
						\label{eq:unitarity}
	\sum_{s'}\int d\Gamma' W^{(ss')}_{\boldsymbol{p}\boldsymbol{p}^{\prime}} =\sum_{s'}\int d\Gamma' W^{(s's)}_{\boldsymbol{p}^{\prime}\boldsymbol{p}} .
	\end{equation}
	As a result, only the deviation $\delta n^{(s)}_{\boldsymbol{p}}$ from equilibrium distribution frunction is present in Eq.~(\ref{Eq:SKCI}).

	The collision integral, Eq.~(\ref{Eq:SKCI}), is set equal to the kinetic terms on the left hand side of Boltzmann equation, Eq.~(\ref{Eq:BE}). With a small temperature gradient in $y$-direction, the linear response approximation reads:
	\begin{equation}
	-\frac{\epsilon^{(s)}_{\boldsymbol{p}}}{T}v^{(s)}_{y,\boldsymbol{p}}\nabla_yT\frac{\partial n^0}{\partial \epsilon}=I^{\mathrm{skew},(s)}_{\boldsymbol{p}}[\delta n ^{(\pm)}].
	\label{Eq:SBE}
	\end{equation}
	If  the scattering rates are known, this equation may be solved for the linear deviations $\delta n ^{(s)}_{\boldsymbol{p}}$. The thermal conductivity is found then from calculating the heat current:
	\begin{equation}
	\label{eq:thermal-current}
	\begin{split}
	&\boldsymbol{j}^{\mathrm{heat,skew}}=\sum_{s=\pm}\int d\Gamma\  \boldsymbol{v}^{(s)}_{\boldsymbol{p}}\epsilon^{(s)}_{\boldsymbol{p}}\delta n^{(s)}_{\boldsymbol{p}},\\
	&j^{\mathrm{heat,skew}}_x=\kappa^{\mathrm{skew}}_{xy}\nabla_yT,\ \ \ \ \ j^{\mathrm{heat,skew}}_y=\kappa^{\mathrm{skew}}_{yy}\nabla_yT, 
	\end{split}
	\end{equation}
	where $\kappa^{\mathrm{skew}}_{xy}$ and $\kappa^{\mathrm{skew}}_{yy}$ are the thermal Hall and longitudinal thermal conductivity respectively.

	\subsection{Side Jump}
	
	The side jump refers to the shift of the center of a Bloch wavepacket upon an impurity scattering event\cite{PhysRevB.73.075318}: $\delta\boldsymbol{r}^{\mathrm{sj},(s^{\prime}s)}_{\boldsymbol{p}^{\prime}\boldsymbol{p}}$, where a quasiparticle is scattered from state $(s,\boldsymbol{p})$ to $(s^{\prime},\boldsymbol{p}^{\prime})$. This leads to the two effects: (i) the quasiparticles acquire an additional velocity and as a result, there is an additional current denoted as $\boldsymbol{j}^{\mathrm{sj}}$; (ii) the distribution function acquires an anomalous part $g^{\mathrm{a},(s)}_{\boldsymbol{j}}$ and the corresponding correction to the current is denoted as $\boldsymbol{j}^{\mathrm{adist}}$.
	While treating the side jump, we disregard the small  skewness of the scattering rates, i.e. we assume here $\delta_Q=0$. In other words we consider the scattering rates to be detailed balance related: $W^{\mathrm{S},(ss')}_{\boldsymbol{p}\boldsymbol{p}^{\prime}}=W^{\mathrm{S},(s's)}_{\boldsymbol{p}^{\prime}\boldsymbol{p}}=W^{(ss')}_{\boldsymbol{p}\boldsymbol{p}^{\prime}}\rvert_{\delta_Q=0}$. The error we commit by this approximation is of the higher order in $m\Delta/k_F$ and $\delta_Q$.

	The  jump velocity is given by:
	\begin{equation}
	\boldsymbol{v}^{\mathrm{sj},(s)}_{\boldsymbol{p}}=\sum_{s^{\prime}}\int d\Gamma^{\prime}W^{\mathrm{S},(s^{\prime}s)}_{\boldsymbol{p}^{\prime}\boldsymbol{p}}\delta\boldsymbol{r}^{\mathrm{sj},(s^{\prime}s)}_{\boldsymbol{p}^{\prime}\boldsymbol{p}}\delta(\epsilon^{(s^{\prime})}_{\boldsymbol{p}^{\prime}}-\epsilon^{(s)}_{\boldsymbol{p}}),
	\end{equation}
	while the corresponding contribution to the transversal current is given by:
	\begin{equation}
	\begin{split}
	j^{\mathrm{sj}}_x=\kappa^{\mathrm{sj}}_{xy}\ \nabla_yT
	=\sum_{s}\int d\Gamma\ v^{\mathrm{sj},(s)}_{x,\boldsymbol{p}}\epsilon^{(s)}_{\boldsymbol{p}}\delta n^{(s)}_{\boldsymbol{p}}
	\end{split}
	\label{Eq:SJCurrent}.
	\end{equation}
The  anomalous distribution $g^{\mathrm{a},(s)}_{\boldsymbol{p}}$ originates  from the collision integral, Eq.~(\ref{Eq:CI}). Notice that the latter is nonlocal in space. It can be transformed into a local form by expanding the distribution function in terms of the side jump displacement:
	\begin{equation}
	n^{(s^{\prime})}_{\boldsymbol{p}^{\prime}}(\boldsymbol{r}^{\prime})=n^{(s^{\prime})}_{\boldsymbol{p}^{\prime}}(\boldsymbol{r})+\delta\boldsymbol{r}^{\mathrm{sj},(s^{\prime}s)}_{\boldsymbol{p}^{\prime}\boldsymbol{p}}\cdot \frac{\partial n^{(s^{\prime})}_{\boldsymbol{p}^{\prime}}(\boldsymbol{r})}{\partial\boldsymbol{r}}
	\end{equation} 
The anomalous distribution $g^{\mathrm{a},(s)}_{\boldsymbol{p}}$ serves to compensating the nonlocality,  by fulfilling the condition:
	\begin{equation}
	\begin{split}
	0=&\sum_{s^{\prime}=\pm}\int d\Gamma^{\prime}
	\delta(\epsilon^{(s^{\prime})}_{\boldsymbol{p}^{\prime}}-\epsilon^{(s)}_{\boldsymbol{p}}) W^{\mathrm{S},(s^{\prime}s)}_{\boldsymbol{p}^{\prime}\boldsymbol{p}}\\
	&\times[g^{\mathrm{a},(s^{\prime})}_{\boldsymbol{p}^{\prime}}+\delta\boldsymbol{r}^{\mathrm{sj},(s^{\prime}s)}_{\boldsymbol{p}^{\prime}\boldsymbol{p}}\cdot \frac{\partial n^{(s^{\prime})}_{\boldsymbol{p}^{\prime}}(\boldsymbol{r})}{\partial\boldsymbol{r}}-g^{\mathrm{a},(s)}_{\boldsymbol{p}}].
	\end{split}
	\label{Eq:AdistDefOriginal}
	\end{equation}
This can be further simplified employing  the linear response relation:
	\begin{equation}
	\frac{\partial n^{(s^{\prime})}_{\boldsymbol{p}^{\prime}}(\boldsymbol{r})}{\partial\boldsymbol{r}}=\frac{\partial T}{\partial\boldsymbol{r}}\frac{\partial n^0(\epsilon^{(s^{\prime})}_{\boldsymbol{p}^{\prime}})}{\partial T}.
	\end{equation}
 Equation~(\ref{Eq:AdistDefOriginal}) can be thus cast into the form:
	\begin{equation}
	\frac{\epsilon^{(s)}_{\boldsymbol{p}}}{T}v^{\mathrm{sj},(s)}_{y,\boldsymbol{p}}\nabla_yT\frac{\partial n^0}{\partial \epsilon}=I^{\mathrm{adist},(s)}_{\boldsymbol{p}}[g^{\mathrm{a},(\pm)}],
	\label{Eq:AdistEqu}
	\end{equation}
	where the collision integral is given below with a symmetric scattering rate:
	\begin{equation}
	\begin{split}
	&I^{\mathrm{adist},(s)}_{\boldsymbol{p}}[g^{\mathrm{a},(\pm)}]=\\
	&\sum_{s^{\prime}=\pm}\int d\Gamma^{\prime}
	W^{\mathrm{S},(s^{\prime}s)}_{\boldsymbol{p}^{\prime}\boldsymbol{p}}\ [g^{\mathrm{a},(s^{\prime})}_{\boldsymbol{p}^{\prime}}-g^{\mathrm{a},(s)}_{\boldsymbol{p}}]\delta(\epsilon^{(s^{\prime})}_{\boldsymbol{p}^{\prime}}-\epsilon^{(s)}_{\boldsymbol{p}}).
	\end{split}
	\end{equation}
	Solving this equation for the `anomalous distribution' $g^{\mathrm{a},(s)}_{\boldsymbol{p}}$, one obtains the second contribution to the transversal current: 
		\begin{equation}
	\begin{split}
	j^{\mathrm{adist}}_x=\kappa^{\mathrm{adist}}_{xy}\nabla_yT=\sum_{s}\int d\Gamma\  v^{(s)}_{x,\boldsymbol{p}}\ \epsilon^{(s)}_{\boldsymbol{p}}\ g^{\mathrm{a},(s)}_{\boldsymbol{p}}.
	\end{split}
	\label{Eq:AdistCurrent}
	\end{equation}
	
An important observation is that the side jump contribution to the thermal Hall conductance is independent of mean free time $\tau$. This can be seen by noticing that the side jump current 
$	j^{\mathrm{Side\ Jump}}_x\propto v^{\mathrm{sj}}\tau$, while 	the side jump velocity is proportional to the scattering rate: $v^{\mathrm{sj}}\propto \tau^{-1}$.  Moreover, as we show below, the side jump contribution to the thermal Hall conductance is of the same order as the intrinsic part.

	\section{Results and discussion}
	\label{sec:results}	
	
With the general framework formulated above we solved the Boltzmann equation in the linear response approximation and evaluated the corresponding thermal conductivites. Details of these calculations are delegated to the 
Appendices. Here we present the  results.

	\begin{figure}[tb]
		\centering
		\includegraphics[width=\linewidth]{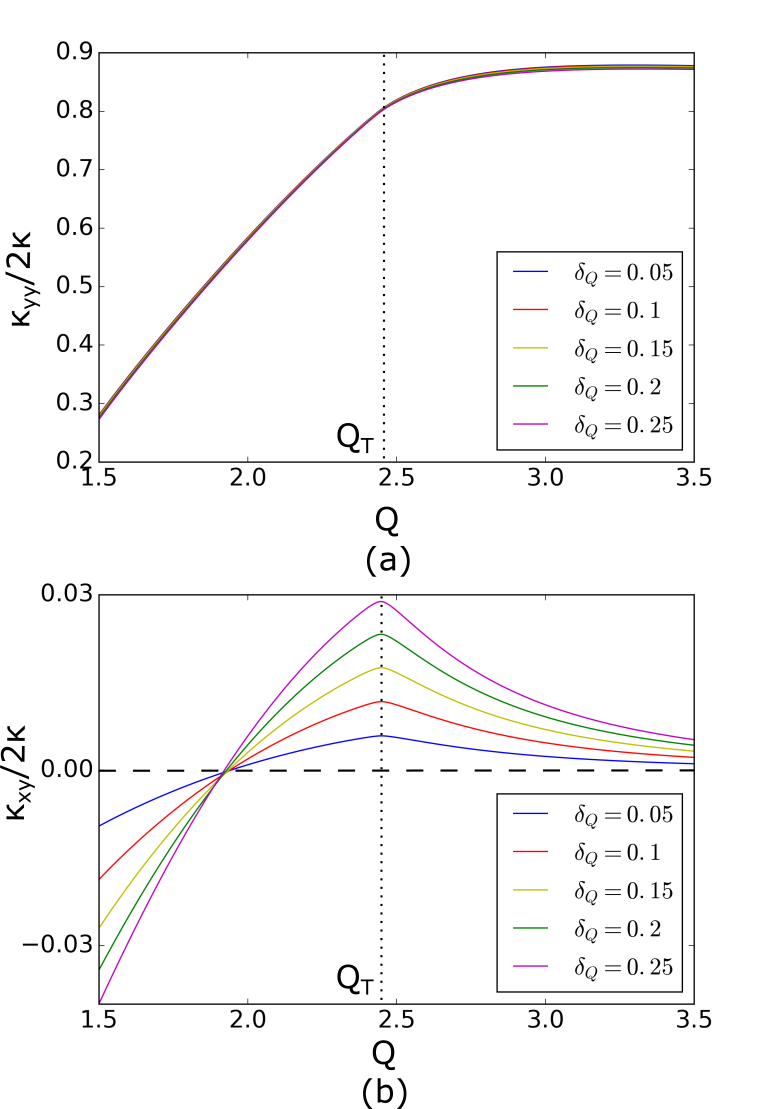}
		\caption{(a) Longitudinal thermal conductivity $\kappa^{\mathrm{skew}}_{yy}$; (b) Thermal Hall conductivity $\kappa^{\mathrm{skew}}_{xy}$ around topological phase transition $Q_T$. Different colors correspond to different impurity strengths. Calculations are done with the same parameters  as in Fig.\ref{fig:BP}. Curves cross zero due to cancellation of two  time reversal symmetry breaking mechanisms: the $p+ip$ superconductivity and the applied supercurrent $Q$. 
		} 
		\label{fig:Cond}
	\end{figure}

	Figure~\ref{fig:Cond}  summaries our  results  for  the skew scattering contribution to the diagonal and off-diagonal thermal conductivity. They are measured in units of  (twice) the thermal conductivity of the  normal metal ($\Delta=0$):
	\begin{equation}
	2\kappa=\frac{\pi^2}{3}\nu Dk_B^2T = 2 g \kappa_Q, 
	\end{equation}
	Here  $D=v_F^2\tau/2$ is the diffusion constant (with $v_F=k_F/m=\sqrt{2\mu/m}$ and $\tau = (2\pi\rho_i \nu V_0^2)^{-1}$ being the normal metal Fermi velocity and relaxation time respectively), $g=h\nu D$ is dimensionless 2D normal state conductance and $\kappa_Q=\pi k_B^2 T/(12\hbar)$ is  thermal conductance quanta.  The topological phase transition takes place at $Q_T=2mv_F$. We present the results 
	for $Q>Q_L=4m\Delta$ only, since in the opposite limit the system is a thermal insulator and the quasiparticle contribution is exponentially small.

	One notices that the diagonal conductivity $\kappa^{\mathrm{skew}}_{yy}$ is essentially  independent on the asymmetry factor $\delta_Q$ and approaches the normal metal value above the topological transition. 
	The off-diagonal   $\kappa^{\mathrm{skew}}_{xy}$ exhibits approximately linear dependence on $\delta_Q$, which in turn is linear in the weak disorder amplitude, $V_0$, Eq.~(\ref{eq:deltaQ}). This is due to the fact that the off-diagonal part, originating from the skew scattering, is proportional to the antisymmetric part of the rate matrix, Eq.~(\ref{eq:antisymmetric}).
The off-diagonal conductivity exhibits maximum at the topological phase transition $Q=Q_T$ and slowly approaches zero deep into the non-topological metallic phase, $Q>Q_T$. Notice that: (i)  $\kappa^{\mathrm{skew}}_{xy}$ is a continuous function of $Q$ through the topological transition and (ii) its magnitude is much less than that of the diagonal $\kappa^{\mathrm{skew}}_{yy}$. The latter observation follows from the fact  that 
	\begin{equation}
	\kappa_{xy}^{\mathrm{skew}}\approx2\kappa \frac{\delta W^\mathrm{antisym}}{{W^\mathrm{S}}}
	\sim g\frac{\Delta^2}{v_F^2}\kappa_Q \sin 2\delta_Q , 
	\label{Eq:EtHC}
	\end{equation}
	where ${W^{\mathrm{S}}}$ is the symmetric part of the scattering rates and we employ  Eq.(\ref{Eq:WS}). In the limit $\Delta/v_F\ll 1$, assumed hereafter, it follows that $\kappa_{xy}\ll \kappa_{yy}$. 
	
	The latter fact is important for the possibility to detect the topological transition within the metallic phase. Indeed, the kinetic theory developed here is valid in the weak scattering limit, when normal-state dimensionless conductance is large $g=h \nu D \gg 1$. Consequently the diagonal thermal conductivity $\kappa^{\mathrm{skew}}_{yy}\propto g\kappa_Q \gg\kappa_Q$. However the off-diagonal one appears to be of the order $\kappa^{\mathrm{skew}}_{xy}\propto g (\Delta/v_F)^2\kappa_Q$.  This value may be comparable to (or even smaller than) the thermal conductance quantum $\kappa_Q$, even in a good metal, $g\gg 1$.
The skew scattering contribution to the diagonal conductivity, $\kappa_{yy}$, on the other hand, is much larger than all other  terms. We thus do not discuss other contributions to the diagonal conductivity.

	\begin{figure}[tb]
		\centering
		\includegraphics[width=\linewidth]{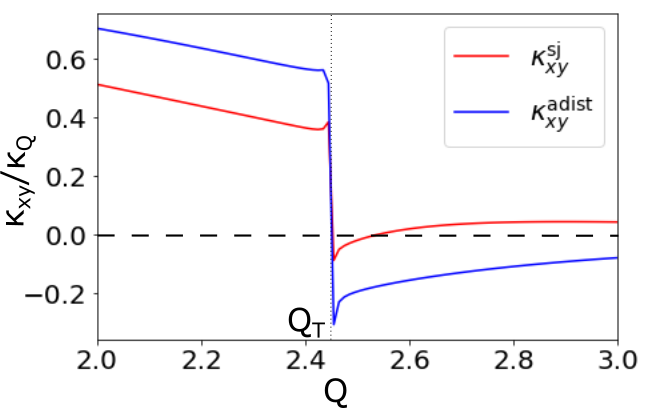}
		\caption{The side jump contribution to the off-diagonal conductance $\kappa_{xy}$. Here $v_F/\Delta =4.9$. (The apparent discrepancy with Eq.~(\ref{Eq:SideJumpJump}) is understood as the consequence of higher order terms in $\Delta/v_F$, see Eq.~(\ref{Eq:SJCondDIs}) in Appendix \ref{sec:SideJump} and the detailed derivation in Appendix \ref{app:SJCal}.)}
		\label{fig:SideJump}
	\end{figure}

Figure~\ref{fig:SideJump} shows  the side jump contributions to the off-diagonal thermal conductivity $\kappa_{xy}$. It exhibits a discontinuity at the topological phase transition. The magnitude of the discontinuity is given by (see Appendix \ref{sec:SideJump}-\ref{app:SJCal}):   
	\begin{equation}
	\Delta\kappa^{\mathrm{sj}}_{xy}\approx\frac{32}{\pi}\frac{\Delta}{v_F}\kappa_Q;  \quad \quad
	\Delta\kappa^{\mathrm{adist}}_{xy}\approx\frac{16}{\pi}\frac{\Delta}{v_F}\kappa_Q. 
	\label{Eq:SideJumpJump}
	\end{equation}
	Those are of the same sign and order of magnitude as the discontinuity in the intrinsic part $\Delta \kappa^{\mathrm{int}}_{xy}=\frac{2\Delta}{v_F}\kappa_Q$. In addition, they should be added together. Therefore, the total discontinuity at the topological phase transition $Q=Q_T$ is:
	\begin{equation}
	\Delta\kappa_{xy}=\left(2+\frac{16}{\pi}+\frac{32}{\pi}\right)\frac{\Delta}{v_F}\kappa_Q.
	\end{equation}
The discontinuity is non-quantized and depends on band-structure parameters, such as $\Delta/v_F$ in our model. On the other hand, it does {\em not} depend on impurities concentration and strength, as long as the sample dimensions exceed elastic mean free path. In the opposite limit of a ballistic sample, the discontinuity is smaller and is given solely by the intrinsic contribution.   	
Comparing the discontinuity with the smooth skew scattering part (\ref{Eq:EtHC}), we observe that the two may be comparable in a wide range of the relevant dimensionless parameters, such as the  ratio $\Delta/v_F$, the normal state conductance $g$ and asymmetry factor $\delta_Q$.

	\begin{figure}[tb]
		\centering
		\includegraphics[width=\linewidth]{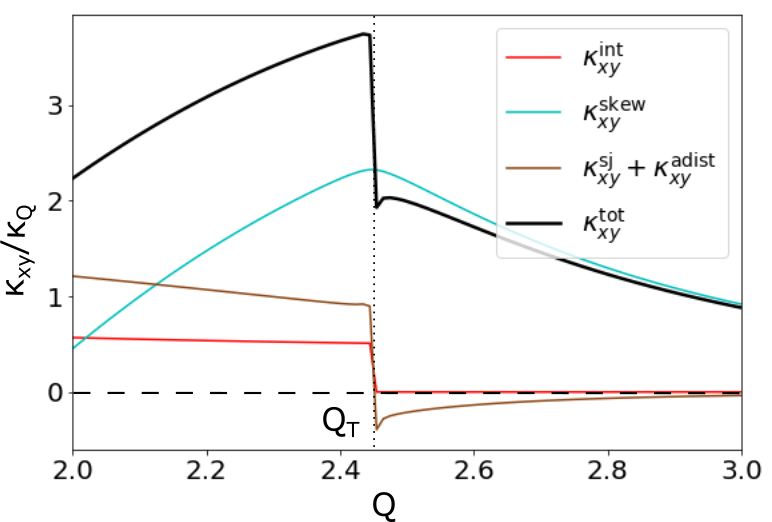}
		\caption{The total thermal Hall conductivity $\kappa_{xy}$ versus supercurrent strength $Q$ (black); the skew scattering contribution (blue); the Berry phase intrinsic contribution  (red); the side jump contribution (brown). The parameters are $g=100$, $v_F/\Delta =4.9$ and $\delta_Q=0.2$.}
		\label{fig:Tot}
	\end{figure}

Figure \ref{fig:Tot} summarises the total off-diagonal conductivity, along with its three constitutive components:  the skew-scattering contribution, Eq.~(\ref{Eq:EtHC}),  the intrinsic contribution Eq.~(\ref{eq:intrinsic}) and side jump, Eq.~(\ref{Eq:SideJumpJump}).  The total $\kappa_{xy}$  exhibits discontinuous jump at the topological phase transition in the limit of small temperature. This discontinuity takes place in the presence of the metallic bulk states and  bulk disorder, as long as the particle-hole symmetry, Eq.~(\ref{eq:particle-hole}), is preserved. Therefore the topological  transition exhibits the sharp signature in transport measurements not only in gapped phases, but also in gapless systems with finite density of states at the Fermi level.  Unlike the gapped case, the corresponding jump is non-quantized.

	One may be concerned if impurity scattering between the bulk and the edge states may change these conclusions. Such processes are not accounted for  in the developed kinetic approach. Indeed, the intrinsic part knows only about the equilibrium distribution function, Eq.~(\ref{eq:intrinsic}), not affected by the impurity scattering. Formally this is due to the fact that the anomalous velocity  $ \boldsymbol{v}^{\mathrm{A}}_{\boldsymbol{p}}=\dot{\boldsymbol{p}}\times\boldsymbol{\Omega}(\boldsymbol{p})$ is already proportional to $\dot{\boldsymbol{p}}\propto \nabla_yT$ and thus to the linear response accuracy one should take the unperturbed distribution function. Physically, it reflects the fact that the edges are in local equilibrium with the adjacent bulk heat reservoir. Since the edge states are exponentially confined to the sample's boundaries, Fig.~\ref{fig:setup}, they can't be affected by the perturbation $\nabla_yT$, which is spread across the bulk.  As a result, the impurity scattering between an edge and an adjacent bulk is exactly the same as in equilibrium: it does not affect the occupation of the edge state and does not change the corresponding thermal current. This reasoning breaks down very close to the topological transition, where the edge states spread into the bulk and thus acquire sensitivity to the perturbation $\nabla_yT$. This happens in earnest only for $|Q-Q_T|\lesssim L_y^{-1/\nu}$, where $\nu$ is the localization length critical exponent.   This leads to a finite size smearing of the discontinuity, which is not present in the thermodynamic limit $L_y\to \infty$. The discontinuity is smeared at a finite temperature, much like the integer quantum Hall transitions.

\section{Acknowledgments}  
	
	We are grateful to A. Andreev, M. Ye, D. Chichinadze, A. Levchenko and M. Sammon for useful discussions. We are also grateful to a referee for his/her insistence to include consideration of the side jump effect.
	This work was supported by NSF Grant No. DMR-1608238.

	\appendix
	
		
	\section{Scattering rates}
	\label{sec:scattering}  
	
	We turn now to the microscopic evaluation of the scattering rates $W^{(ss^{\prime})}_{\boldsymbol{p}\boldsymbol{p}^{\prime}}$. The off-diagonal response, $\kappa_{xy}$, is a consequence of the asymmetry in the scattering rate, Eq.~(\ref{eq:not-symmetry}), which appears in the absence of the time-reversal symmetry and is known as the {\em skew} scattering mechanism. In our model the time reversal symmetry is already broken by $p+ip$ superconductivity, Eq.~(\ref{CH}),  and the  supercurrent $Q$ further introduces anisotropy to the system. As a result, the detailed balance is violated, which leads to anomalous transport in SPT metal. 	
	
	We restrict ourselves with the weak point-like impurities with impurity density $\rho_i$. Therefore the scattering rates are given by:
	\begin{equation}
		W^{(ss^{\prime})}_{\boldsymbol{p}\boldsymbol{p}^{\prime}}=2\pi\rho_i|T^{(ss^{\prime})}_{\boldsymbol{p}\boldsymbol{p}^{\prime}}|^2
		\label{Eq:Wdefinition}
	\end{equation}
	where $T^{(ss^{\prime})}_{\boldsymbol{p}\boldsymbol{p}^{\prime}}$ is the scattering amplitudes of an individual impurity between {\em quasiparticle} states.
	
	Scattering process should contain all orders in Born series and the Lippmann-Schwinger equation is used to find scattering $T$-matrix $\hat{T}$, which determines scattering amplitudes\cite{PhysRevB.92.100506,0038-5670-27-11-A06,sinitsyn2007semiclassical}. Consider a generic Hamiltonian:
	\begin{equation}
		\hat{H}=\hat{H}_0 + \hat{V}, 
	\end{equation}
	where $\hat{H}_0$ is the Hamiltonian free of impurities and $\hat{V}$ is the impurity potential. The scattering $T$-matrix is formally defined by the Lippmann-Schwinger equation\cite{altland2010condensed}:   
	\begin{equation}
		\hat{T}=\hat{V}+\hat{V}\ \hat{G}_0(i\omega)\ \hat{T},
		\label{Eq:DT}
	\end{equation}
	where $\hat{G}_0(i\omega)=(i\omega-\hat{H}_0)^{-1}$ is the bare Green's function. The scattering amplitudes $T^{(ss^{\prime})}_{\boldsymbol{p}\boldsymbol{p}^{\prime}}$ between quasiparticle states could be determined from $T$-matrix:
	\begin{equation}
		T^{(ss^{\prime})}_{\boldsymbol{p}\boldsymbol{p}^{\prime}}=\langle \boldsymbol{p},s|\hat{T}|\boldsymbol{p}^{\prime},s^{\prime}\rangle, 
	\end{equation}
	where $|\boldsymbol{p},s=\pm\rangle$ is the quasiparticle eigenstate.

	Following Li, Andreev and Spivak\cite{PhysRevB.92.100506} we first solve Eq.~(\ref{Eq:DT}) in the original particle-hole basis and   then transform to the quasiparticle basis as:
	\begin{equation}
		T^{(ss^{\prime})}_{\boldsymbol{p}\boldsymbol{p}^{\prime}}=\sum_{a,b=p,h}\langle \boldsymbol{p},s|\boldsymbol{p},a\rangle\langle \boldsymbol{p},a|\hat{T}|\boldsymbol{p}^{\prime},b\rangle\langle \boldsymbol{p}^{\prime},b|\boldsymbol{p}^{\prime},s^{\prime}\rangle.
		\label{Eq:TPHtQP}
	\end{equation}
	A remark about notations is due:  the labels $a(b)=p,h$ are labeling  particle-hole states, while  $s(s^{\prime})=\pm$ are labeling the quasiparticle bands. Thus, the summation indices $a,b$ is over particle-hole states and $\psi_{\boldsymbol{p},a}^{(s=\pm)}=\langle \boldsymbol{p},a|\boldsymbol{p},s=\pm\rangle$ are the elements of the quasiparticle wavefunctions, given by Eq.~(\ref{Eq:QPWF}). 
	In the  particle-hole basis  equation (\ref{Eq:DT})  takes the form\cite{PhysRevB.92.100506}:
	\begin{equation}
		T^{(ab)}_{\boldsymbol{p}\boldsymbol{p}^{\prime}}=V^{(ab)}_{\boldsymbol{p}\boldsymbol{p}^{\prime}} + \sum_{\boldsymbol{k}}\sum_{c,d = p,h}V^{(ac)}_{\boldsymbol{p}\boldsymbol{k}}G^{(cd)}_0(i\omega,\boldsymbol{k})T^{(db)}_{\boldsymbol{k}\boldsymbol{p}^{\prime}},
		\label{Eq:LSEPH}
	\end{equation}
	where $V^{(ab)}_{\boldsymbol{p}\boldsymbol{p}^{\prime}}=\langle \boldsymbol{p},a|\hat{V}|\boldsymbol{p}^{\prime},b\rangle$ and $G^{(ab)}_0(i\omega,\boldsymbol{k})=\langle \boldsymbol{k},a|\hat{G}_0(i\omega)|\boldsymbol{k},b\rangle$.
	In this same basis the Hamiltonian $\hat{H}_0$ takes the form of the BdG Hamiltonian (\ref{Eq:Ham}) and the point-like impurity potential $\hat{V}$ is independent of momenta  $V^{(ab)}_{\boldsymbol{p}\boldsymbol{p}^{\prime}}=V_0\sigma_z^{\mathrm{a}b}$.  Therefore, the scattering matrix $T^{(ab)}_{\boldsymbol{p}\boldsymbol{p}^{\prime}}$ is independent of momenta and is only a function of the frequency $i\omega$. Hence Eq.(\ref{Eq:LSEPH}) can be rewritten in a compact way  as a matrix equation in the  particle-hole space\cite{PhysRevB.92.100506}:
	\begin{equation}
		\hat T(i\omega)=V_0\sigma_z+V_0\sigma_z\sum_{\boldsymbol{k}}\hat G_0(i\omega,\boldsymbol{k})\hat T(i\omega),
		\label{LSE}
	\end{equation}
	where the Green function is defined as $\hat G_0(i\omega, \boldsymbol{k})=(i\omega-H_{\mathrm{BdG}}(\boldsymbol{k}))^{-1}$. The formal solution for the scattering matrix $\hat T(i\omega)$ is:
	\begin{equation}
		\hat T(i\omega)=[1-V_0\sigma_z\sum_{\boldsymbol{k}}\hat G_0(i\omega,\boldsymbol{k})]^{-1}\ V_0\sigma_z.
		\label{Eq:FormalSOlT}
	\end{equation}
	
	Close to the  topological phase transition, $|Q-Q_T|\ll8m\Delta$, the summation of the Green function over momenta  $\sum_{\boldsymbol{k}}\hat G_0(i\omega,\boldsymbol{k})$ may be separated into two parts: $\sum_{|\boldsymbol{k}|<4m\Delta}$ and $\sum_{|\boldsymbol{k}|>4m\Delta}$, as indicated in Fig.~\ref{fig:FSS}. The Hamiltonian, Eq.(\ref{Eq:Ham}), has very different asymptotic behavior in these two regimes. Since we considered small temperatures,  only the quasiparticle states near the zero energy are important. Therefore, the quantity of interest here is actually $
	\sum_{\boldsymbol{k}}\hat G_0(i\omega\rightarrow 0,\boldsymbol{k})$. 
	For small momenta, $|\boldsymbol{k}|<4m\Delta$, the Hamiltonian in Eq.(\ref{Eq:Ham}) is approximately a tilted massive Dirac Hamiltonian with the mass of $M_Q=\frac{1}{2}\frac{k_F}{m}(Q-Q_T)$:
	\begin{equation}
		H_{\mathrm{BdG}}(\boldsymbol{k})\approx Qk_x/2m
		-2\Delta k_y\sigma_x+2\Delta k_x\sigma_y+M_Q\sigma_z. 
		\label{Eq:TiltedDirac}
	\end{equation}
	with corrections on the order of $\mathcal{O}(\boldsymbol{k}^2/m)$ and $\mathcal{O}((Q-Q_T)^2/m)$.
	Hence,  summation of the Green function over momenta is:
	\begin{equation}
		\begin{split}
			&\sum_{|\boldsymbol{k}|<4m\Delta}\hat G_0(i\omega\rightarrow 0,\boldsymbol{k})\\
			&= \sum_{|\boldsymbol{k}|<4m\Delta}\frac{Qk_x/2m
				-2\Delta k_y\sigma_x+2\Delta k_x\sigma_y+M_Q\sigma_z}{(Qk_x/2m)^2
				-4\Delta^2 \boldsymbol{k}^2-M_Q^2}.
		\end{split}
	\end{equation}
	It's clear that the terms with numerator linear in momenta $k_x$ and $k_y$ vanish upon summation. Hence the result is proportional to the mass term $M_Q\sigma_z$. Performing the integration, as detailed in Appendix \ref{app:integral}, one finds: 
	\begin{equation}
		\sum_{|\boldsymbol{k}|<4m\Delta}\hat G_0(i\omega\rightarrow 0,\boldsymbol{k})=-\frac{m\sigma_z}{4\pi k_F}(Q-Q_T).
		\label{Eq:SK}
	\end{equation}
	For large momenta $|\boldsymbol{k}|>4m\Delta$, one can neglect the off-diagonal terms in the Hamiltonian (\ref{CH}), leaving  two copies of the normal metal decoupled from each other. 

	Therefore,  summation of the Green results in the 2D density of states of a metal:
	\begin{equation}
		\sum_{|\boldsymbol{k}|>4m\Delta}G_0(i\omega\rightarrow 0,\boldsymbol{k})\approx i\pi \nu\sigma_0, 
		\label{Eq:LK}
	\end{equation}
	where $\nu$ is a density of states at the chemical potential. Taken together Eqs.~(\ref{Eq:SK}) and (\ref{Eq:LK}) yield:
	\begin{equation}
		\sum_{\boldsymbol{k}}\hat G_0(i\omega\rightarrow 0,\boldsymbol{k})\approx-\frac{m\sigma_z}{4\pi k_F}(Q-Q_T)+i\pi \nu. 
	\end{equation}

	The scattering matrix $\hat T(i\omega\rightarrow 0)$ can be obtained  from Eq.~(\ref{Eq:FormalSOlT})\cite{PhysRevB.92.100506}:
	\begin{equation}
		\label{eq:Tparticle-hole}
		\hat T(i\omega\rightarrow0)=\begin{pmatrix}
			f(Q)e^{i\delta_Q} & 0\\
			0 & -f(Q)e^{-i\delta_Q}
		\end{pmatrix}, 
	\end{equation}
	where
	\begin{equation}
		\label{eq:fQ}
		f(Q)e^{i\delta_Q}=\frac{V_0}{1+\frac{mV_0}{4\pi k_F}(Q-Q_T)-i\pi\nu V_0}. 
	\end{equation}
	It's clear that the scattering matrix $\hat T$ is a smooth function of the supercurrent  $Q$ around the topological phase transition at $Q=Q_T$. As explained below, the phase $\delta_Q$  is responsible for the skew scattering. Close to the topological phase transition one finds from Eq.~(\ref{eq:fQ}): 
	\begin{equation}
		\tan \delta_Q\approx\pi \nu V_0, 
		\label{Eq:DA}
	\end{equation}
	approximately a constant proportional to the impurity potential $V_0$.

	To find scattering amplitudes between the quasiparticle states one needs to transform the particle-hole matrix (\ref{eq:Tparticle-hole}) to the quasiparticle basis according to Eq.~(\ref{Eq:TPHtQP}): 
	\begin{equation}
		\label{eq:TtoT}
		T^{(ss^{\prime})}_{\boldsymbol{p}\boldsymbol{p}^{\prime}}=\left(\psi^{(s)}_{\boldsymbol{p}}\right)^\dagger \hat T(i\omega\rightarrow0) \psi^{(s^{\prime})}_{\boldsymbol{p}^{\prime}},
	\end{equation}
	where the quasiparticle spinors are given by Eq.~(\ref{Eq:QPWF}). 		The scattering rates are defined as $W^{(ss^{\prime})}_{\boldsymbol{p}\boldsymbol{p}^{\prime}}=2\pi \rho_i|T^{(ss^{\prime})}_{\boldsymbol{p}\boldsymbol{p}^{\prime}}|^2$, where $\rho_i$ is impurity density. A straightforward calculation outlined in Appendix \ref{app:rates} leads to the following structure of the rate matrix: 
	\begin{equation}
		W^{(ss^{\prime})}_{\boldsymbol{p}\boldsymbol{p}^{\prime}}=W^{0,(ss^{\prime})}_{\boldsymbol{p}\boldsymbol{p}^{\prime}}+W^{1,(ss^{\prime})}_{\boldsymbol{p}\boldsymbol{p}^{\prime}}\cos(\theta_{\boldsymbol{p}^{\prime}}-\theta_{\boldsymbol{p}}-2\delta_Q), 
		\label{Eq:WS}
	\end{equation}
	where $W^{0,(ss^{\prime})}_{\boldsymbol{p}\boldsymbol{p}^{\prime}}$ and $W^{1,(ss^{\prime})}_{\boldsymbol{p}\boldsymbol{p}^{\prime}}$ are symmetric with respect to the momenta $\boldsymbol{p}\leftrightarrow \boldsymbol{p}^{\prime}$ interchange and  $\theta_{\boldsymbol{p}}$ and $\theta_{\boldsymbol{p}^{\prime}}$ are the azimuthal angles of the corresponding momenta (i.e. $\exp(i\theta_{\boldsymbol{p}})=(p_x+ip_y)/|\boldsymbol{p}|$). As can be seen from Eq.~(\ref{eq:A5}), the dependence of the rates on the azimuthal angles comes solely from  $p+ip$  superconductivity present in the Hamiltonian (\ref{TBH}). This in turn leads to an {\em antisymmetric} part of the rate matrix	
	\begin{equation}
		\label{eq:antisymmetric} 
		\delta W_\mathrm{antisym} = W^1\sin 2\delta_Q\sin(\theta_{\boldsymbol{p}^{\prime}}-\theta_{\boldsymbol{p}}),  
	\end{equation}
	which is responsible for  the off-diagonal thermal conductivity.

	Once the scattering rates $W^{(ss^{\prime})}_{\boldsymbol{p}\boldsymbol{p}^{\prime}}$ are found (see Appendix \ref{app:rates}), the linearized Boltzmann equation 	(\ref{Eq:SBE}) may be solved numerically for the 
	deviation $\delta n^{(s)}_{\boldsymbol{p}}$ of the distribution function from its equilibrium value. It is convenient to parametrize such deviations in terms of angles labeling position along the two, $(\pm)$, Fermi surfaces, see inset in   
	Fig.~\ref{fig:FSS}, 
	\begin{equation}
		\delta n^{(s)}_{\boldsymbol{p}} = \sum_{n=1}^{N_\mathrm{cutoff} }[a^{(s)}_n\cos(n\varphi)+b^{(s)}_n\sin(n\varphi)], 
	\end{equation}
	where $N_\mathrm{cutoff}$ is a suitably chosen cutoff (typically a few dozens). The absence of the $n=0$  term  reflects the  elastic nature of the scattering, which ensures  quasiparticle number conservation at any energy. 
	We solve for the harmonic amplitudes $a_n^{(s)}$ and $b_n^{(s)}$ and then evaluate linear response current according to Eq.~(\ref{eq:thermal-current}). 
	
	One can do the same for solving for the `anomalous distribution' $g^{\mathrm{a},(s)}_{\boldsymbol{p}}$, once the side jump velocity is found in the Appendix \ref{sec:SideJump}-\ref{app:SJCal}.

	\section{Calculations leading to Eq.(\ref{Eq:SK})}
	\label{app:integral}

	In this section, we show details for deriving Eq.(\ref{Eq:SK}):
	\begin{equation}
	\sum_{|\boldsymbol{k}|<4m\Delta}\hat G_0(i\omega\rightarrow 0,\boldsymbol{k})=-\frac{m\sigma_z}{4\pi k_F}(Q-Q_T).
	\end{equation}
	The first step is to replace summation by integration:
	\begin{equation}
	\sum_{|\boldsymbol{k}|<4m\Delta}\rightarrow\int_{|\boldsymbol{k}|<4m\Delta}\frac{d^2k}{(2\pi)^2}
	\end{equation}

	Notice the Green's function takes the following form at small momenta:
	\begin{equation}
	\begin{split}
	G(i\omega\rightarrow0,\boldsymbol{k})=\frac{Qk_x/2m
		-2\Delta k_y\sigma_x+2\Delta k_x\sigma_y+M_Q\sigma_z}{(Qk_x/2m)^2
		-4\Delta^2 \boldsymbol{k}^2-M_Q^2}.
	\end{split}
	\end{equation}
	The terms in the numerator here linear in momenta vanishes upon integration. Thus, the quantity to calculate is:
\begin{widetext}	
	\begin{equation}
		\begin{split}
		\int_{|\boldsymbol{k}|<4m\Delta}\frac{d^2k}{(2\pi)^2}G(i\omega\rightarrow0,\boldsymbol{k})=\int_{|\boldsymbol{k}|<4m\Delta}\frac{d^2k}{(2\pi)^2}\frac{M_Q\sigma_z}{(Qk_x/2m)^2
			-4\Delta^2 \boldsymbol{k}^2-M_Q^2}.
		\end{split}	
	\end{equation}
The angular integration results in:
	\begin{equation}
	\int_{0}^{2\pi}d\theta\frac{M_Q\sigma_z}{(Qk/2m)^2\cos^2\theta
		-4\Delta^2 k^2-M_Q^2}=-\frac{2\pi M_Q\sigma_z}{\sqrt{M_Q^2+4\Delta^2k^2}\sqrt{M_Q^2+4\Delta^2k^2-\frac{Q^2}{4m^2}k^2}}\theta(M_Q^2+4\Delta^2k^2-\frac{Q^2}{4m^2}k^2)
	\end{equation}
	
	where $\theta(\dots)$ is the step function. Its presence  puts an additional constraint on the integration over the magnitude of momentum $k$. To simplify further calculations we focus on the vicinity of the topological phase transition $|Q-Q_T|\ll8m\Delta$ and thus $Q^2/4m^2\approx v_F^2$, while $\Delta\ll v_F$. The above integral thus simplifies down to:
	\begin{equation}
	-\!\!\!\int\limits_{k<4m\Delta}\!\!\frac{kdk}{(2\pi)^2}\frac{2\pi M_Q\sigma_z\theta(M_Q^2-v_F^2k^2)}{\sqrt{M_Q^2}\sqrt{M_Q^2-v_F^2k^2}} \approx -\frac{M_Q\sigma_z}{2\pi v_F^2}.
	\end{equation}
Noticing that the `mass' $M_Q=\frac{1}{2}\frac{k_F}{m}(Q-Q_T)$, where $v_F=k_F/m$, one arrives at Eq.(\ref{Eq:SK}).

	\section{Scattering Rates}
	\label{app:rates}
	
	Here we derive explicit  expressions for the scattering rates $W^{(ss^{\prime})}_{\boldsymbol{p}\boldsymbol{p}^{\prime}}$ in terms of $\boldsymbol{d}(\boldsymbol{p})$ functions in the BdG Hamiltonian Eq.~(\ref{Eq:Ham}). As discussed in the main text one needs to transform the scattering $T$-matrix $\hat{T}(i\omega\rightarrow0)$, Eq.~(\ref{eq:Tparticle-hole}), in the particle-hole basis to the  quasiparticle basis with the help of the  wavefunctions $\psi^{(s)}_{\boldsymbol{p}}$ through Eq.(\ref{Eq:Wdefinition}):
	\begin{equation}
	W^{(ss^{\prime})}_{\boldsymbol{p}\boldsymbol{p}^{\prime}}=2\pi\rho_i|T^{(ss^{\prime})}_{\boldsymbol{p}\boldsymbol{p}^{\prime}}|^2
	\end{equation}
	where according to Eq.(\ref{eq:TtoT}):
	\begin{equation}
	T^{(ss^{\prime})}_{\boldsymbol{p}\boldsymbol{p}^{\prime}}=(\psi^{(s)}_{\boldsymbol{p}})^{\dagger}\hat{T}(i\omega\rightarrow0)\psi^{(s^{\prime})}_{\boldsymbol{p}^{\prime}},
	\end{equation}
where the spinors may be written in terms of  $\mathbf{d}(\boldsymbol{p})$ and the azimuthal angle $\theta_{\boldsymbol{p}}$ as:
	\begin{equation}
	\begin{split}
	&\psi^{(+)}_{\boldsymbol{p}}=\mathcal{N}_{\boldsymbol{p}}\begin{pmatrix}
	|\boldsymbol{d}(\boldsymbol{p})|+d_z(\boldsymbol{p})\\
	2i\Delta pe^{i\theta_{\boldsymbol{p}}}
	\end{pmatrix};\\
	&\psi^{(-)}_{\boldsymbol{p}}=\mathcal{N}_{\boldsymbol{p}}\begin{pmatrix}
	2i\Delta pe^{-i\theta_{\boldsymbol{p}}}\\
	|\boldsymbol{d}(\boldsymbol{p})|+d_z(\boldsymbol{p})
	\end{pmatrix}.
	\end{split}
	\label{Eq:QPWF1}
	\end{equation}
	This particular form of wavefunctions is a direct result of $p$-wave superconductivity and leads to a nontrivial angular dependence of $W^{(ss^{\prime})}_{\boldsymbol{p}\boldsymbol{p}^{\prime}}$. The scattering matrix in particle-hole space is:
	\begin{equation}
	\hat{T}(i\omega\rightarrow0)=\begin{pmatrix}
	f(Q)e^{i\delta_Q} & 0\\
	0 & -f(Q)e^{-i\delta_Q}
	\end{pmatrix}.
	\end{equation} 
		
This way one finds for  the scattering amplitudes between the quasiparticle states:
	\begin{equation}
							\label{eq:A5}
	\begin{split}
	&T^{(++)}_{\boldsymbol{p}\boldsymbol{p}^{\prime}}=\mathcal{N}_{\boldsymbol{p}}\mathcal{N}_{\boldsymbol{p}^{\prime}}
	\begin{pmatrix}
	|\boldsymbol{d}(\boldsymbol{p})|+d_z(\boldsymbol{p}), & -2i\Delta pe^{-i\theta_{\boldsymbol{p}}}
	\end{pmatrix}
	\begin{pmatrix}
	f(Q)e^{i\delta_Q} & 0\\
	0 & -f(Q)e^{-i\delta_Q}
	\end{pmatrix}
	\begin{pmatrix}
	|\boldsymbol{d}(\boldsymbol{p}^{\prime})|+d_z(\boldsymbol{p}^{\prime})\\
	2i\Delta pe^{i\theta_{\boldsymbol{p}^{\prime}}}
	\end{pmatrix}\\
	&\ \ \ \ \ \ \ \  =\mathcal{N}_{\boldsymbol{p}}\mathcal{N}_{\boldsymbol{p}^{\prime}}[(|\boldsymbol{d}(\boldsymbol{p})|+d_z(\boldsymbol{p}))(|\boldsymbol{d}(\boldsymbol{p}^{\prime})|+d_z(\boldsymbol{p}^{\prime}))f(Q)e^{i\delta_Q}-4\Delta^2pp^{\prime}f(Q)e^{i(\theta_{\boldsymbol{p}^{\prime}}-\theta_{\boldsymbol{p}}-\delta_Q)}]\\
	&\ \ \ \ \ \ \ \  =\mathcal{N}_{\boldsymbol{p}}\mathcal{N}_{\boldsymbol{p}^{\prime}}f(Q)e^{i\delta_Q}[(|\boldsymbol{d}(\boldsymbol{p})|+d_z(\boldsymbol{p}))(|\boldsymbol{d}(\boldsymbol{p}^{\prime})|+d_z(\boldsymbol{p}^{\prime}))-4\Delta^2pp^{\prime}e^{i(\theta_{\boldsymbol{p}^{\prime}}-\theta_{\boldsymbol{p}}-2\delta_Q)}];\\
	&T^{(--)}_{\boldsymbol{p}\boldsymbol{p}^{\prime}}=-\mathcal{N}_{\boldsymbol{p}}\mathcal{N}_{\boldsymbol{p}^{\prime}}f(Q)e^{-i\delta_Q}[(|\boldsymbol{d}(\boldsymbol{p})|+d_z(\boldsymbol{p}))(|\boldsymbol{d}(\boldsymbol{p}^{\prime})|+d_z(\boldsymbol{p}^{\prime}))-4\Delta^2pp^{\prime}e^{-i(\theta_{\boldsymbol{p}^{\prime}}-\theta_{\boldsymbol{p}}-2\delta_Q)}];\\
	&T^{(+-)}_{\boldsymbol{p}\boldsymbol{p}^{\prime}}=i\mathcal{N}_{\boldsymbol{p}}\mathcal{N}_{\boldsymbol{p}^{\prime}}f(Q)e^{i(\delta_Q-\theta_{\boldsymbol{p}^{\prime}})}[2\Delta p^{\prime}(|\boldsymbol{d}(\boldsymbol{p})|+d_z(\boldsymbol{p}))+2\Delta p(|\boldsymbol{d}(\boldsymbol{p}^{\prime})|+d_z(\boldsymbol{p}^{\prime}))e^{i(\theta_{\boldsymbol{p}^{\prime}}-\theta_{\boldsymbol{p}}-2\delta_Q)}];\\
	&T^{(-+)}_{\boldsymbol{p}\boldsymbol{p}^{\prime}}=-i\mathcal{N}_{\boldsymbol{p}}\mathcal{N}_{\boldsymbol{p}^{\prime}}f(Q)e^{i(\delta_Q+\theta_{\boldsymbol{p}})}[2\Delta p(|\boldsymbol{d}(\boldsymbol{p}^{\prime})|+d_z(\boldsymbol{p}^{\prime}))+2\Delta p^{\prime}(|\boldsymbol{d}(\boldsymbol{p})|+d_z(\boldsymbol{p}))e^{i(\theta_{\boldsymbol{p}^{\prime}}-\theta_{\boldsymbol{p}}-2\delta_Q)}]. 
	\end{split}
	\end{equation}
The scattering rates are then found with the help of Eq.(\ref{Eq:Wdefinition}):
	\begin{equation}
	\begin{split}
	W^{(++)}_{\boldsymbol{p}\boldsymbol{p}^{\prime}}&=2\pi\rho_i|T^{(++)}_{\boldsymbol{p}\boldsymbol{p}^{\prime}}|^2\\
	& =2\pi\rho_if^2(Q)\mathcal{N}^2_{\boldsymbol{p}}\mathcal{N}^2_{\boldsymbol{p}^{\prime}}\\
	&\times\{[(|\boldsymbol{d}(\boldsymbol{p})|+d_z(\boldsymbol{p}))(|\boldsymbol{d}(\boldsymbol{p}^{\prime})|+d_z(\boldsymbol{p}^{\prime}))-4\Delta^2pp^{\prime}\cos (\theta_{\boldsymbol{p}^{\prime}}-\theta_{\boldsymbol{p}}-2\delta_Q)]^2\\
	&+[4\Delta^2pp^{\prime}\sin (\theta_{\boldsymbol{p}^{\prime}}-\theta_{\boldsymbol{p}}-2\delta_Q)]^2\}\\
	& =2\pi\rho_if^2(Q)\mathcal{N}^2_{\boldsymbol{p}}\mathcal{N}^2_{\boldsymbol{p}^{\prime}}\\
	&\times\{(|\boldsymbol{d}(\boldsymbol{p})|+d_z(\boldsymbol{p}))^2(|\boldsymbol{d}(\boldsymbol{p}^{\prime})|+d_z(\boldsymbol{p}^{\prime}))^2+16\Delta^4p^2p^{\prime\ 2}\\
	&-8\Delta^2pp^{\prime}(|\boldsymbol{d}(\boldsymbol{p})|+d_z(\boldsymbol{p}))(|\boldsymbol{d}(\boldsymbol{p}^{\prime})|+d_z(\boldsymbol{p}^{\prime}))\cos (\theta_{\boldsymbol{p}^{\prime}}-\theta_{\boldsymbol{p}}-2\delta_Q)\};\\
	W^{(--)}_{\boldsymbol{p}\boldsymbol{p}^{\prime}}&=2\pi\rho_if^2(Q)\mathcal{N}^2_{\boldsymbol{p}}\mathcal{N}^2_{\boldsymbol{p}^{\prime}}\\
	&\times\{(|\boldsymbol{d}(\boldsymbol{p})|+d_z(\boldsymbol{p}))^2(|\boldsymbol{d}(\boldsymbol{p}^{\prime})|+d_z(\boldsymbol{p}^{\prime}))^2+16\Delta^4p^2p^{\prime\ 2}\\
	&-8\Delta^2pp^{\prime}(|\boldsymbol{d}(\boldsymbol{p})|+d_z(\boldsymbol{p}))(|\boldsymbol{d}(\boldsymbol{p}^{\prime})|+d_z(\boldsymbol{p}^{\prime}))\cos (\theta_{\boldsymbol{p}^{\prime}}-\theta_{\boldsymbol{p}}-2\delta_Q)\};\\
	W^{(+-)}_{\boldsymbol{p}\boldsymbol{p}^{\prime}}&=2\pi\rho_if^2(Q)\mathcal{N}^2_{\boldsymbol{p}}\mathcal{N}^2_{\boldsymbol{p}^{\prime}};\\
	&\times\{4\Delta^2p^2(|\boldsymbol{d}(\boldsymbol{p}^{\prime})|+d_z(\boldsymbol{p}^{\prime}))^2+4\Delta^2p^{\prime2}(|\boldsymbol{d}(\boldsymbol{p})|+d_z(\boldsymbol{p}))^2\\
	&+8\Delta^2pp^{\prime}(|\boldsymbol{d}(\boldsymbol{p})|+d_z(\boldsymbol{p}))(|\boldsymbol{d}(\boldsymbol{p}^{\prime})|+d_z(\boldsymbol{p}^{\prime}))\cos (\theta_{\boldsymbol{p}^{\prime}}-\theta_{\boldsymbol{p}}-2\delta_Q)\};\\
	W^{(-+)}_{\boldsymbol{p}\boldsymbol{p}^{\prime}}&=2\pi\rho_if^2(Q)\mathcal{N}^2_{\boldsymbol{p}}\mathcal{N}^2_{\boldsymbol{p}^{\prime}}\\
	&\times\{4\Delta^2p^2(|\boldsymbol{d}(\boldsymbol{p}^{\prime})|+d_z(\boldsymbol{p}^{\prime}))^2+4\Delta^2p^{\prime2}(|\boldsymbol{d}(\boldsymbol{p})|+d_z(\boldsymbol{p}))^2\\
	&+8\Delta^2pp^{\prime}(|\boldsymbol{d}(\boldsymbol{p})|+d_z(\boldsymbol{p}))(|\boldsymbol{d}(\boldsymbol{p}^{\prime})|+d_z(\boldsymbol{p}^{\prime}))\cos (\theta_{\boldsymbol{p}^{\prime}}-\theta_{\boldsymbol{p}}-2\delta_Q)\}. 
	\end{split}
	\end{equation}
	
Although cumbersome, these expressions allow to estimate the magnitude of thermal Hall conductivity, $\kappa_{xy}/2\kappa$, resulting in Eq.(\ref{Eq:EtHC}). Indeed, the typical symmetric and antisymmetric parts of $W^{(ss^{\prime})}_{\boldsymbol{p}\boldsymbol{p}^{\prime}}$ have the form:
	\begin{equation}
	\begin{split}
	&W^0=2\pi\rho_if^2(Q)\mathcal{N}^2_{\boldsymbol{p}}\mathcal{N}^2_{\boldsymbol{p}^{\prime}}\times\{(|\boldsymbol{d}(\boldsymbol{p})|+d_z(\boldsymbol{p}))^2(|\boldsymbol{d}(\boldsymbol{p}^{\prime})|+d_z(\boldsymbol{p}^{\prime}))^2+16\Delta^4p^2p^{\prime\ 2}\};\\
	&W^1=2\pi\rho_if^2(Q)\mathcal{N}^2_{\boldsymbol{p}}\mathcal{N}^2_{\boldsymbol{p}^{\prime}}\times8\Delta^2pp^{\prime}(|\boldsymbol{d}(\boldsymbol{p})|+d_z(\boldsymbol{p}))(|\boldsymbol{d}(\boldsymbol{p}^{\prime})|+d_z(\boldsymbol{p}^{\prime})).
	\end{split}
	\end{equation}
	Assuming $\Delta\ll v_F$ and  $Q\approx Q_T$, one finds:
	\begin{equation}
	\frac{\kappa_{xy}}{2\kappa}\approx\frac{\Delta^2 pp^{\prime}}{(p^2/m)(p^{\prime2}/m)}\sin 2\delta_Q\approx\frac{\Delta^2}{v_F^2}\sin 2\delta_Q.
	\end{equation}
	
	\end{widetext}

	\section{Side Jump Displacement and Velocity}
	\label{sec:SideJump}
	
	\subsection{Side Jump Displacement and Velocity}
	
	In this section, we turn to the microscopic calculation of the side jump displacement $\delta\boldsymbol{r}^{\mathrm{sj},(s^{\prime}s)}_{\boldsymbol{p}^{\prime}\boldsymbol{p}}$ and the side jump velocity $\boldsymbol{v}^{\mathrm{sj},(s)}_{\boldsymbol{p}}$. The actual calculation is rather tedious. Instead of providing all the details, we simply sketch the derivation process. Then we present the most relevant terms in the side jump velocity and estimate the discontinuity in the off-diagonal conductance around topological phase transition.
	
	We employed the gauge invariant definition of the side jump displacement derived in Ref.~[\onlinecite{PhysRevB.73.075318}]:
	\begin{equation}
		\begin{split}
			\delta\boldsymbol{r}^{\mathrm{sj},(s^{\prime}s)}_{\boldsymbol{p}^{\prime}\boldsymbol{p}}=&\left(\psi^{(s^{\prime})}_{\boldsymbol{p}^{\prime}}\right)^\dagger i\partial_{\boldsymbol{p}^{\prime}} \psi^{(s^{\prime})}_{\boldsymbol{p}^{\prime}}-\left(\psi^{(s)}_{\boldsymbol{p}}\right)^\dagger i\partial_{\boldsymbol{p}} \psi^{(s)}_{\boldsymbol{p}}\\
			&-D_{\boldsymbol{p},\boldsymbol{p}^{\prime}} \left[\arg \left(T^{(s^{\prime}s)}_{\boldsymbol{p}^{\prime}\boldsymbol{p}}\right)\right]
		\end{split}
	\end{equation}
	where the differential operator is defined as $D_{\boldsymbol{p},\boldsymbol{p}^{\prime}}=\partial_{\boldsymbol{p}}+\partial_{\boldsymbol{p}^{\prime}}$.
	The terms in the first line is also known as the Berry connection, $\mathcal{A}^{(s)}_{\boldsymbol{p}}=\left(\psi^{(s)}_{\boldsymbol{p}}\right)^\dagger i\partial_{\boldsymbol{p}} \psi^{(s)}_{\boldsymbol{p}}$. Given the quasiparticle wavefunctions Eq.~(\ref{Eq:QPWF}), it's straightforward to show:
	\begin{equation}
		\mathcal{A}^{(s)}_{\boldsymbol{p}}=s\ \mathcal{N}^2_{\boldsymbol{p}}\left[d_x(\boldsymbol{p})\partial_{\boldsymbol{p}}d_y(\boldsymbol{p})-d_y(\boldsymbol{p})\partial_{\boldsymbol{p}}d_x(\boldsymbol{p})\right]
	\end{equation}
	As before, $s=\pm$ is the quasiparticle band index.
	
	The second line involves the scattering amplitude $T^{(s^{\prime}s)}_{\boldsymbol{p}^{\prime}\boldsymbol{p}}$, which is derived in the previous Appendix \ref{app:rates}. Here, we make further simplifications. First, as argued in Sec.~\ref{sec:kinetic}, we took $\delta_Q=0$ when the side jump is under consideration. This is the same as neglecting the interplay between the skew scattering and the side jump effect. Second, we focus ourselves to the regime very close to the topological phase transition, $Q\approx Q_T$. With those two simplifications, the scattering amplitudes take a simple form following Eq.~(\ref{eq:TtoT}):
	\begin{equation}
		T^{(s^{\prime}s)}_{\boldsymbol{p}^{\prime}\boldsymbol{p}}=V_0\left(\psi^{(s^{\prime})}_{\boldsymbol{p}^{\prime}}\right)^\dagger\sigma_z\psi^{(s)}_{\boldsymbol{p}}
	\end{equation}
	The impurity potential $V_0$ does not depend on the momenta. As a result:
	\begin{equation}
		\arg \left(T^{(s^{\prime}s)}_{\boldsymbol{p}^{\prime}\boldsymbol{p}}\right)=\arg \left[\left(\psi^{(s^{\prime})}_{\boldsymbol{p}^{\prime}}\right)^\dagger\sigma_z\psi^{(s)}_{\boldsymbol{p}}\right]
	\end{equation}
	Or equivalently:
	\begin{equation}
		\arg \left(T^{(s^{\prime}s)}_{\boldsymbol{p}^{\prime}\boldsymbol{p}}\right)=\arg \left[\left(\tilde{\psi}^{(s^{\prime})}_{\boldsymbol{p}^{\prime}}\right)^\dagger\sigma_z\tilde{\psi}^{(s)}_{\boldsymbol{p}}\right]
	\end{equation}
	where $\tilde{\psi}^{(s)}_{\boldsymbol{p}}=\psi^{(s)}_{\boldsymbol{p}}/\mathcal{N}_{\boldsymbol{p}}$ is the unnormalized wavefunction. The unnormalized wavefunction $\tilde{\psi}^{(s)}_{\boldsymbol{p}}$ simplifies the detailed calculations a little bit.
	Taking the derivative on the argument of a complex valued function (in this case $T^{(s^{\prime}s)}_{\boldsymbol{p}^{\prime}\boldsymbol{p}}$ or $\left(\tilde{\psi}^{(s^{\prime})}_{\boldsymbol{p}^{\prime}}\right)^\dagger\sigma_z\tilde{\psi}^{(s)}_{\boldsymbol{p}}$) gives:
	\begin{equation}
	\resizebox{0.5\textwidth}{!}{
		$\begin{split}
		D_{\boldsymbol{p},\boldsymbol{p}^{\prime}} \left[\arg \left(T^{(s^{\prime}s)}_{\boldsymbol{p}^{\prime}\boldsymbol{p}}\right)\right]=\frac{\Im \left[\left(\tilde{\psi}^{(s)}_{\boldsymbol{p}}\right)^\dagger\sigma_z\tilde{\psi}^{(s^{\prime})}_{\boldsymbol{p}^{\prime}}D_{\boldsymbol{p},\boldsymbol{p}^{\prime}}\left[\left(\tilde{\psi}^{(s^{\prime})}_{\boldsymbol{p}^{\prime}}\right)^\dagger\sigma_z\tilde{\psi}^{(s)}_{\boldsymbol{p}}\right]\right]}{\left|\left(\tilde{\psi}^{(s^{\prime})}_{\boldsymbol{p}^{\prime}}\right)^\dagger\sigma_z\tilde{\psi}^{(s)}_{\boldsymbol{p}}\right|^2}
		\end{split}$}
	\end{equation}
	In the numerator, $\Im[\cdots]$ means taking the imaginary part of the quantity in the square bracket.

	Above we derived the working expressions for the side jump displacement. However, it's more important to define the side jump velocity:
	\begin{equation}
		\boldsymbol{v}^{\mathrm{sj},(s)}_{\boldsymbol{p}}=\sum_{s'}\int\Gamma^{\prime}\ W^{\mathrm{S},(s^{\prime}s)}_{\boldsymbol{p}^{\prime}\boldsymbol{p}}\delta\boldsymbol{r}^{\mathrm{sj},(s^{\prime}s)}_{\boldsymbol{p}^{\prime}\boldsymbol{p}}\delta(\epsilon^{(s^{\prime})}_{\boldsymbol{p}^{\prime}}-\epsilon^{(s)}_{\boldsymbol{p}})
		\label{Eq:SJVel}
	\end{equation}
	Close to the topological phase transition $Q\approx Q_T$, the scattering rates is given by:
	\begin{equation}
		W^{\mathrm{S},(s^{\prime}s)}_{\boldsymbol{p}^{\prime}\boldsymbol{p}}=2\pi\rho_i\lvert V_0\rvert^2\ \mathcal{N}^2_{\boldsymbol{p}}\mathcal{N}^2_{\boldsymbol{p}^{\prime}}\ \lvert\left(\tilde{\psi}^{(s^{\prime})}_{\boldsymbol{p}^{\prime}}\right)^\dagger\sigma_z\tilde{\psi}^{(s)}_{\boldsymbol{p}}\rvert^2
	\end{equation}
	
\begin{figure}[tb]
		\centering
		\includegraphics[width=\linewidth]{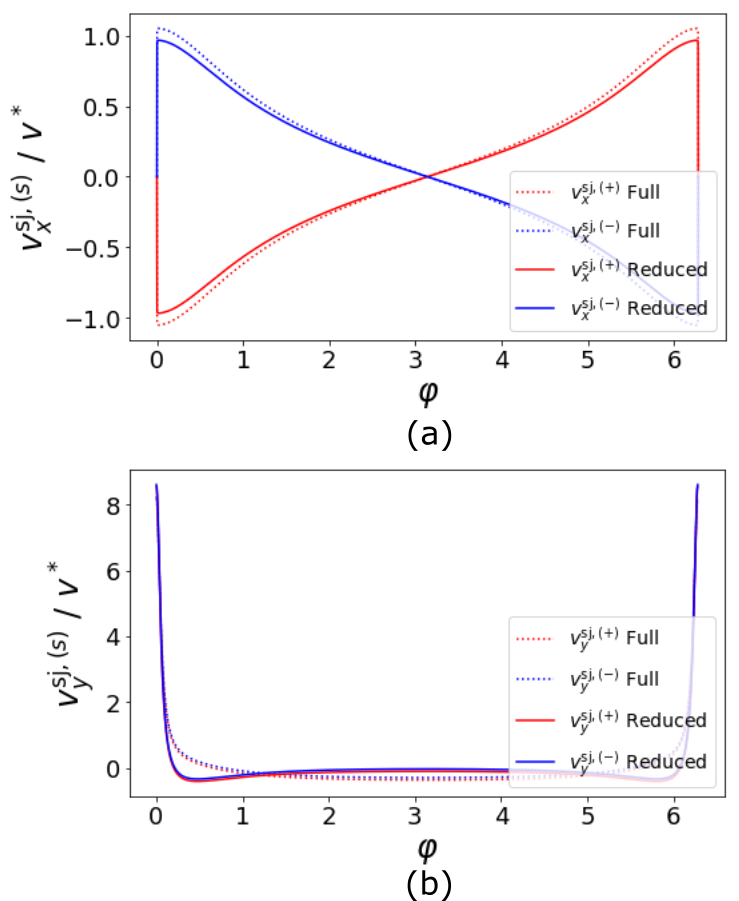}
		\caption{The side jump velocity for states on the Fermi surfaces, parameterized by the angle $\varphi$ as in the inset of Fig.\ref{fig:FSS}. The solid line is plotted from the reduced expression, Eq.~(\ref{Eq:SJVelReduced}), while the dotted line is obtained from the full definition, Eq.~(\ref{Eq:SJVel}). The unit here is chosen to be $v^*=\Delta/(k_Fl)$. Both plots are plotted right before topological phase transition $Q<Q_T$. For (b), the red and blue curves should be identical. For visualization, they were being shifted by a small number.
		} 
		\label{fig:VelSJ}
	\end{figure}

	Following this procedure, one would end up with a rather cumbersome expression. Fortunately, if one focus on the vicinity of topological phase transition, the side jump velocity is governed by part of the full expression:
	\begin{subequations}
		\begin{align}
			&v^{\mathrm{sj},(s)}_{x,\boldsymbol{p}}=s\ \frac{2}{\pi}(1-\frac{\pi}{2}\frac{\Delta}{v_F})\frac{M_Q}{\lvert M_Q\rvert}\frac{v_F p_y}{\lvert d(\boldsymbol{p})\rvert}  \frac{\Delta}{k_Fl} \label{Eq:SJVelReduced1} \\
			&v^{\mathrm{sj},(s)}_{y,\boldsymbol{p}}=-\Delta k_F\frac{ 1-8\Delta^2p^2\mathcal{N}^2_{\boldsymbol{p}}}{\lvert d(\boldsymbol{p})\rvert}\frac{\Delta}{k_Fl} \label{Eq:SJVelReduced2}
		\end{align}
		\label{Eq:SJVelReduced}
	\end{subequations}
	Fig. \ref{fig:VelSJ} is a comparison of the side jump velocity obtained from the full definition, Eq.~(\ref{Eq:SJVel}), and the reduced expression, Eq.~(\ref{Eq:SJVelReduced}).
	
	Before explaining the features of Eq.~(\ref{Eq:SJVelReduced}), we should remind our readers that the terms in Eq.~(\ref{Eq:SJVelReduced}) are not the only nonvanishing terms for the side jump velocity. However, those are the only terms that change sign over the topological phase transition. Therefore, they should give rise to a discontinuity in the off-diagonal conductance, $\kappa_{xy}$. From now on, we simply focus on Eq.~(\ref{Eq:SJVelReduced}).
	
	In Eq.~(\ref{Eq:SJVelReduced}), both expressions change sign over the topological phase transition. In Eq.~(\ref{Eq:SJVelReduced1}), the side jump velocity in $x$-direction is explicitly proportional to a factor of $M_Q/\lvert M_Q\rvert$, where $M_Q = \frac{1}{2}\frac{k_F}{m}(Q-Q_T)$ is the Dirac mass in Eq.~(\ref{Eq:TiltedDirac}).
	
	Meanwhile, for the side jump velocity in $y$-direction Eq.~(\ref{Eq:SJVelReduced2}), the factor of $1-8\Delta^2p^2\mathcal{N}^2_{\boldsymbol{p}}$ changes sign. The normalization factor is $\mathcal{N}_{\boldsymbol{p}}= \big[2|\boldsymbol{d}(\boldsymbol{p})|(|\boldsymbol{d}(\boldsymbol{p})|+d_z(\boldsymbol{p}))\big]^{-1/2}$. For small momentum $p<\Lambda=4m\Delta$, the quasiparticles are described by Eq.~(\ref{Eq:TiltedDirac}). The side jump velocity takes the following form:
	\begin{equation}
		v^{\mathrm{sj},(s)}_{y,\boldsymbol{p}}=-\frac{\Delta k_F}{\lvert d(\boldsymbol{p})\rvert}\frac{\Delta}{k_Fl}\times\left\{ \begin{array}{ccc}
			\frac{M_ Q}{\lvert M_Q\rvert} & \ \ \ p< \frac{\lvert M_Q\rvert}{2\Delta}\\\ \\
			\frac{M_Q}{2\Delta p} & \ \ \ \frac{\lvert M_Q\rvert}{2\Delta}<p<\Lambda
		\end{array} \right.
	\end{equation}
	
	As a result, the side jump velocity in both $x$- and $y$-directions is at least proportional to the sign of $M_Q$. Thus, they should change sign over the topological phase transition. This directly leads to the discontinuity in the thermal Hall conductance, $\kappa_{xy}$.
	
	The second observation is that both expressions are proportional to the inverse of the mean free path $l=v_F\tau$. $\tau^{-1}=2\pi\rho_i\nu\lvert V_0\rvert^2$ is the mean free time. As discussed, the direct result of this feature is that the side jump contribution to $\kappa_{xy}$ is independent of the impurity strength, if we confine ourselves to the diffusive regime.

	\subsection{Estimation of the Discontinuity in $\kappa_{xy}$}
	
	With the side jump velocity in Eq.~(\ref{Eq:SJVelReduced}), we were able to estimate the discontinuity in the thermal Hall conductance with relaxation time approximation. The correct sign and order of magnitude could be found. Technical details are provided in the Appendix \ref{app:SJCal}.
	
	Start with the discontinuity in $\kappa^{\mathrm{sj}}_{xy}$:
	\begin{equation}
		\Delta\kappa^{\mathrm{sj}}_{xy}\approx\frac{32}{\pi}\frac{\Delta}{v_F}\left(1-(2+\frac{\pi}{2})\frac{\Delta}{v_F}\right)\kappa_Q
		\label{Eq:SJCondDIs}
	\end{equation}
	With relaxation time approximation, Eq.~(\ref{Eq:SBE}) reduces to:
	\begin{equation}
		-\frac{\epsilon^{(s)}_{\boldsymbol{p}}}{T}v^{(s)}_{y,\boldsymbol{p}}\nabla_yT\frac{\partial n^0}{\partial \epsilon}=-\frac{\delta n^{(s)}_{\boldsymbol{p}}}{\tau}
	\end{equation}
	It's straightforward that the distribution function is:
	\begin{equation}
		\delta n^{(s)}_{\boldsymbol{p}}=\frac{\epsilon^{(s)}_{\boldsymbol{p}}}{T}v^{(s)}_{y,\boldsymbol{p}}\tau\nabla_yT\frac{\partial n^0}{\partial \epsilon}
	\end{equation}
	The first part of the side jump current could be defined from Eq.~(\ref{Eq:SJCurrent}):
	\begin{equation}
		j^{\mathrm{sj}}_x=\sum_{s}\int d\Gamma\ v^{\mathrm{sj},(s)}_{x,\boldsymbol{p}}\epsilon^{(s)}_{\boldsymbol{p}}\frac{\epsilon^{(s)}_{\boldsymbol{p}}}{T}v^{(s)}_{y,\boldsymbol{p}}\tau\nabla_yT\frac{\partial n^0}{\partial \epsilon}
	\end{equation}
	Notice the presence of mean free time $\tau$ in the above equation. It will cancel the  $\tau^{-1}$ dependence in the side jump velocity. Thus, the side jump contribution should be independent of mean free time $\tau$ (or mean free path $l$). By performing the integration explicitly as in Appendix \ref{app:SJCal}, one finds:
	\begin{equation}
		j^{\mathrm{sj}}_x=\frac{4}{3}\frac{\Delta}{v_F}\left(1-\frac{2\Delta}{v_F}\right)\left(1-\frac{\pi\Delta}{2v_F}\right)\frac{M_Q}{\lvert M_Q\rvert}T\nabla_yT
	\end{equation}
	The discontinuity in $\kappa^{\mathrm{sj}}_{xy}$ can be read off directly to the second order in $\Delta/v_F$:
	\begin{equation}
		\begin{split}
			\Delta\kappa^{\mathrm{sj}}_{xy}\approx&\frac{8}{3}\frac{\Delta}{v_F}\left(1-(2+\frac{\pi}{2})\frac{\Delta}{v_F}\right)T\\
			=&\frac{32}{\pi}\frac{\Delta}{v_F}\left(1-(2+\frac{\pi}{2})\frac{\Delta}{v_F}\right)\kappa_Q
		\end{split}
	\end{equation}
	In the second line, $\kappa_Q=\frac{\pi}{12}T$ was factored out.
	
	Similarly, the anomalous distribution function can be estimated with the relaxation time approximation:
	\begin{equation}
		g^{\mathrm{a},(s)}_{\boldsymbol{p}}=-\frac{\epsilon^{(s)}_{\boldsymbol{p}}}{T}v^{\mathrm{sj},(s)}_{y,\boldsymbol{p}}\tau\nabla_yT\frac{\partial n^0}{\partial \epsilon}
	\end{equation}
	The associated current is given by Eq.~(\ref{Eq:AdistCurrent}):
	\begin{equation}
		j^{\mathrm{adist}}_x=\sum_{s}\int d\Gamma\  v^{(s)}_{x,\boldsymbol{p}}\ \epsilon^{(s)}_{\boldsymbol{p}}\ g^{\mathrm{a},(s)}_{\boldsymbol{p}}
	\end{equation}
	Following the similar procedure, the discontinuity in $\kappa^{\mathrm{adist}}_{xy}$ was estimated to be:
	\begin{equation}
		\Delta\kappa^{\mathrm{adist}}_{xy}\approx\frac{16}{\pi}\frac{\Delta}{v_F}\kappa_Q
	\end{equation}

	\section{Integration over side jump velocity}
	\label{app:SJCal}
	
	In this appendix, we provide the details for evaluation of the integrals of side jump contribution to the off-diagonal thermal conductance.
	
	Start with $\Delta\kappa^{\mathrm{sj}}_{xy}$:
	\begin{equation}
	\Delta\kappa^{\mathrm{sj}}_{xy}\approx\frac{32}{\pi}\frac{\Delta}{v_F}\left(1-(2+\frac{\pi}{2})\frac{\Delta}{v_F}\right)\kappa_Q
	\end{equation}
	With relaxation time approximation, Eq.~(\ref{Eq:SBE}) reduces to:
	\begin{equation}
	-\frac{\epsilon^{(s)}_{\boldsymbol{p}}}{T}v^{(s)}_{y,\boldsymbol{p}}\nabla_yT\frac{\partial n^0}{\partial \epsilon}=-\frac{\delta n^{(s)}_{\boldsymbol{p}}}{\tau}
	\end{equation}
	It's straightforward that the distribution function is:
	\begin{equation}
	\delta n^{(s)}_{\boldsymbol{p}}=\frac{\epsilon^{(s)}_{\boldsymbol{p}}}{T}v^{(s)}_{y,\boldsymbol{p}}\tau\nabla_yT\frac{\partial n^0}{\partial \epsilon}
	\end{equation}
	The first part of the side jump current could be defined from Eq.~(\ref{Eq:SJCurrent}):
	\begin{equation}
	j^{\mathrm{sj}}_x=\sum_{s}\int d\Gamma\ v^{\mathrm{sj},(s)}_{x,\boldsymbol{p}}\epsilon^{(s)}_{\boldsymbol{p}}\frac{\epsilon^{(s)}_{\boldsymbol{p}}}{T}v^{(s)}_{y,\boldsymbol{p}}\tau\nabla_yT\frac{\partial n^0}{\partial \epsilon}
	\end{equation}
	The trick to perform the integration is to change the integration measure:
	\begin{equation}
	d\Gamma=\frac{dp_xdp_y}{(2\pi)^2}\rightarrow\frac{dp_xd\epsilon}{(2\pi)^2}2\lvert\frac{dp_y}{d\epsilon}\rvert
	\label{Eq:ChangeVAr}
	\end{equation}
	Notice that both the side jump velocity and the group velocity ($v^{(s)}_{y,\boldsymbol{p}}\propto s \times p_y$) are proportional to $p_y$. Therefore the integrand in the current expression is even in $p_y$. One could just focus on the upper plane with $p_y>0$. Focusing on the upper half plane would introduce a factor of $2$ in the last expression in Eq.~(\ref{Eq:ChangeVAr}).
	Then the current expression reduces to 
	\begin{equation}
	j^{\mathrm{sj}}_x=2\sum_{s}\int \frac{dp_xd\epsilon}{(2\pi)^2}\ v^{\mathrm{sj},(s)}_{x,\boldsymbol{p}}\epsilon^{(s)}_{\boldsymbol{p}}\frac{\epsilon^{(s)}_{\boldsymbol{p}}}{T}\tau\nabla_yT\frac{\partial n^0}{\partial \epsilon}
	\end{equation}
	Further simplification can be made if one think about really low temperature. Namely, only a tiny energy window around Fermi surfaces is under consideration:
	\begin{equation}
	j^{\mathrm{sj}}_x=2\sum_{s}\int \frac{dp_x}{(2\pi)^2}\ v^{\mathrm{sj},(s)}_{x,\boldsymbol{p}}\tau\int d\epsilon\frac{\epsilon^2}{T}\nabla_yT\frac{\partial n^0}{\partial \epsilon}
	\end{equation}
	The second integral gives a factor of:
	\begin{equation}
	\int d\epsilon\frac{\epsilon^2}{T}\nabla_yT\frac{\partial n^0}{\partial \epsilon}=-\frac{2\pi^2}{3}T\nabla_yT
	\end{equation}
	The first integral should be understood as an integration on the Fermi surfaces. Namely, the momenta are implicitly constrained by $\epsilon^{(s)}_{\boldsymbol{p}}=0$. The integral could be evaluated analytically:
	\begin{widetext}

	\begin{equation}
	\int \frac{dp_x}{(2\pi)^2}\ v^{\mathrm{sj},(s)}_{x,\boldsymbol{p}}\tau=\frac{2}{(2\pi)^2}\frac{\Delta}{v_F}\left(1-\frac{2\Delta}{v_F}\right)\left(1-\frac{\pi\Delta}{2v_F}\right)\frac{M_Q}{\lvert M_Q\rvert}
	\end{equation}
	
	Details of this integral is as follows:
	\begin{equation}
	\int \frac{dp_x}{(2\pi)^2}\ v^{\mathrm{sj},(s)}_{x,\boldsymbol{p}}\tau=\int \frac{dp_x}{(2\pi)^2}\ \frac{2}{\pi}\left(1-\frac{\pi\Delta}{2v_F}\right)\frac{M_Q}{\lvert M_Q\rvert}\frac{v_F p_y}{\lvert d(\boldsymbol{p})\rvert}\frac{\Delta\tau}{k_Fl}
	\end{equation}
	First, notice that the momenta are constrained by $\epsilon^{(s)}_{\boldsymbol{p}}=0$, or more explicitly:
	\begin{equation}
	v_Fp_x+\sqrt{\frac{p^4}{4m^2}+4\Delta^2p^2}=0
	\end{equation}
	As a result, we could change the integration variable to:
	\begin{equation}
	dp_x=\frac{1}{2v_F}\frac{p^3/m^2+8\Delta^2p}{\sqrt{\frac{p^4}{4m^2}+4\Delta^2p^2}}dp
	\end{equation}
	Meanwhile, $p_y$ could also be expressed in terms of $p$:
	\begin{equation}
	p_y=\sqrt{(1-\frac{4\Delta^2}{v_F^2})p^2-\frac{p^4}{4m^2v_F^2}}
	\end{equation}
	Notice that $\lvert d(\boldsymbol{p})\rvert=\sqrt{\frac{p^4}{4m^2}+4\Delta^2p^2}$. With all those quantities, the integral transforms to:
	\begin{equation}
	\begin{split}
	\int \frac{dp_x}{(2\pi)^2}\ v^{\mathrm{sj},(s)}_{x,\boldsymbol{p}}\tau\ =\ &\frac{2}{\pi}\left(1-\frac{\pi\Delta}{2v_F}\right)\frac{M_Q}{\lvert M_Q\rvert}\frac{\Delta}{2v_Fk_F}\int \frac{dp}{(2\pi)^2}\ \frac{p^3/m^2+8\Delta^2p}{\frac{p^4}{4m^2}+4\Delta^2p^2}\sqrt{(1-\frac{4\Delta^2}{v_F^2})p^2-\frac{p^4}{4m^2v_F^2}}\\
	=\ &\frac{2}{\pi}\left(1-\frac{\pi\Delta}{2v_F}\right)\frac{M_Q}{\lvert M_Q\rvert}\frac{\Delta}{v_F}\int \frac{dx}{(2\pi)^2}\ \frac{4x^2+8\Delta^2/v_F^2}{x^2+4\Delta^2/v_F^2}\sqrt{(1-\frac{4\Delta^2}{v_F^2})-x^2}
	\end{split}
	\end{equation}
	In the last line, $x=p/(2mv_F)$. The integration in the last line could be evaluated analytically:
	\begin{equation}
	\int \frac{dx}{(2\pi)^2}\ \frac{4x^2+8\Delta^2/v_F^2}{x^2+4\Delta^2/v_F^2}\sqrt{(1-\frac{4\Delta^2}{v_F^2})-x^2}=\frac{\pi}{(2\pi)^2}(1-\frac{2\Delta}{v_F})
	\end{equation}

	Therefore, putting everything together, we obtained:
	\begin{equation}
	j^{\mathrm{sj}}_x=-\frac{4}{3}\frac{\Delta}{v_F}\left(1-\frac{2\Delta}{v_F}\right)\left(1-\frac{\pi\Delta}{2v_F}\right)\frac{M_Q}{\lvert M_Q\rvert}T\nabla_yT
	\end{equation}
	where the summation over the band index $\sum_s [\cdots]$ gives an additional factor of $2$. The discontinuity in $\kappa_{xy}$ can be read off directly to the second order in $\Delta/v_F$:
	\begin{equation}
	\begin{split}
	\Delta\kappa^{\mathrm{sj}}_{xy}\approx&\frac{8}{3}\frac{\Delta}{v_F}\left(1-(2+\frac{\pi}{2})\frac{\Delta}{v_F}\right)T\\
	=&\frac{32}{\pi}\frac{\Delta}{v_F}\left(1-(2+\frac{\pi}{2})\frac{\Delta}{v_F}\right)\kappa_Q
	\end{split}
	\end{equation}
	In the second line, $\kappa_Q=\frac{\pi}{12}T$ was factored out.
	
	Similarly, the anomalous distribution function can be estimated from Eq.~(\ref{Eq:AdistEqu}) with the relaxation time approximation:
	\begin{equation}
	g^{\mathrm{a},(s)}_{\boldsymbol{p}}=-\frac{\epsilon^{(s)}_{\boldsymbol{p}}}{T}v^{\mathrm{sj},(s)}_{y,\boldsymbol{p}}\tau\nabla_yT\frac{\partial n^0}{\partial \epsilon}
	\end{equation}
	The associated current is given by Eq.~(\ref{Eq:AdistCurrent}):
	\begin{equation}
	j^{\mathrm{adist}}_x=\sum_{s}\int d\Gamma\  v^{(s)}_{x,\boldsymbol{p}}\ \epsilon^{(s)}_{\boldsymbol{p}}\ g^{\mathrm{a},(s)}_{\boldsymbol{p}}
	\end{equation}
	Following the similar procedure, we arrived at the expression:
	\begin{equation}
	j^{\mathrm{adist}}_x=-2\sum_{s}\int_{p_y>0} \frac{dp_y}{(2\pi)^2}\ v^{\mathrm{sj},(s)}_{y,\boldsymbol{p}}\tau\int d\epsilon\frac{\epsilon^2}{T}\nabla_yT\frac{\partial n^0}{\partial \epsilon}
	\end{equation}
	As before,
	\begin{equation}
	\int d\epsilon\frac{\epsilon^2}{T}\nabla_yT\frac{\partial n^0}{\partial \epsilon}=-\frac{2\pi^2}{3}T\nabla_yT
	\end{equation}
	The other integral can be calculates as follows:
	\begin{equation}
	\int_{p_y>0} \frac{dp_y}{(2\pi)^2}\ v^{\mathrm{sj},(s)}_{y,\boldsymbol{p}}\tau
	\end{equation}
	Keep in mind that the momenta are constrained by the condition $\epsilon^{(s)}_{\boldsymbol{p}}=0$. We focus on small momentum regime, where the side jump velocity takes a large value:
	\begin{equation}
	\int_{p<\Lambda} \frac{dp_y}{(2\pi)^2}\ v^{\mathrm{sj},(s)}_{y,\boldsymbol{p}}\tau
	\end{equation}
	where $\Lambda=4m\Delta$ is the cutoff when the quasiparticles ceased to behave like a Dirac particle. The above integration could be evaluated in two steps. First:
	
	\begin{equation}
	\begin{split}
	\int_{p<\frac{\lvert M_Q\rvert}{2\Delta}} \frac{dp_y}{(2\pi)^2}\ v^{\mathrm{sj},(s)}_{y,\boldsymbol{p}}\tau=-\int_{p<\frac{\lvert M_Q\rvert}{2\Delta}} \frac{dp_y}{(2\pi)^2}\ \frac{\Delta k_F}{\lvert d(\boldsymbol{p})\rvert}\frac{\Delta\tau}{k_Fl}\frac{M_Q}{\lvert M_Q\rvert}
	\end{split}
	\end{equation}
	Notice that the `exact' expression for $\lvert d(\boldsymbol{p})\rvert=\sqrt{4\Delta^2p^2+M_Q^2}$. For small momentum $p<\frac{\lvert M_Q\rvert}{2\Delta}$, we could simplify $\lvert d(\boldsymbol{p})\rvert\approx \lvert M_Q\rvert$. At the same time, $p_y<\frac{\lvert M_Q\rvert}{2\Delta}$ has approximately the same limitation. With this simplification, the integration is straightforward:
	\begin{equation}
	\int_{p_y<\frac{\lvert M_Q\rvert}{2\Delta}} \frac{dp_y}{(2\pi)^2}\ v^{\mathrm{sj},(s)}_{y,\boldsymbol{p}}\tau=-\frac{1}{(2\pi)^2}\frac{\Delta}{2v_F}\frac{M_Q}{\lvert M_Q\rvert}
	\end{equation}
	Second, we evaluate:
	\begin{equation}
	\int_{\frac{\lvert M_Q\rvert}{2\Delta}<p<\Lambda} \frac{dp_y}{(2\pi)^2}\ v^{\mathrm{sj},(s)}_{y,\boldsymbol{p}}\tau=-\int_{\frac{\lvert M_Q\rvert}{2\Delta}<p<\Lambda} \frac{dp_y}{(2\pi)^2}\frac{\Delta k_F}{\lvert d(\boldsymbol{p}\rvert)}\frac{\Delta\tau}{k_Fl}\frac{M_Q}{2\Delta p}
	\end{equation}
	For momentum in the region $\frac{\lvert M_Q\rvert}{2\Delta}<p<\Lambda$, one direct simplification is that $\lvert d(\boldsymbol{p})\rvert=2\Delta p$. Namely, $M_Q$ is neglected. With this simplification, the constraint $\epsilon^{(s)}_{\boldsymbol{p}}=0$ reads:
	\begin{equation}
	\begin{split}
	v_Fp_x+s2\Delta p=0\rightarrow p_y=\sqrt{1-\frac{4\Delta^2}{v_F^2}}p\approx p
	\end{split}
	\end{equation}
	The integration reduces to
	\begin{equation}
	\begin{split}
	\int_{\frac{\lvert M_Q\rvert}{2\Delta}<p<\Lambda} \frac{dp_y}{(2\pi)^2}\ v^{\mathrm{sj},(s)}_{y,\boldsymbol{p}}\tau=-\int_{\frac{\lvert M_Q\rvert}{2\Delta}<p<\Lambda} \frac{dp}{(2\pi)^2}\frac{M_Q}{4 v_Fp^2}
	=-\frac{1}{(2\pi)^2}\frac{M_Q}{4v_F}\left(-\frac{1}{\Lambda}+\frac{2\Delta}{\lvert M_Q\rvert}\right)
	\approx-\frac{1}{(2\pi)^2}\frac{M_Q}{\lvert M_Q\rvert}\frac{\Delta}{2v_F}
	\end{split}
	\end{equation}
	Putting together the two-step integration, we get:
	\begin{equation}
	\int_{p<\Lambda} \frac{dp_y}{(2\pi)^2}\ v^{\mathrm{sj},(s)}_{y,\boldsymbol{p}}\tau=-\frac{1}{(2\pi)^2}\frac{M_Q}{\lvert M_Q\rvert}\frac{\Delta}{v_F}
	\end{equation}
	With the results for the integrations, one finds the current to be:
	\begin{equation}
	j^{\mathrm{adist}}_x=-\frac{2}{3}\frac{\Delta}{v_F}\frac{M_Q}{\lvert M_Q\rvert}T\nabla_yT
	\end{equation}
	The discontinuity in $\kappa_{xy}$ was estimated to be:
	\begin{equation}
	\begin{split}
	\Delta\kappa^{\mathrm{adist}}_{xy}\approx\frac{4}{3}\frac{\Delta}{v_F}T=\frac{16}{\pi}\frac{\Delta}{v_F}\kappa_Q
	\end{split}
	\end{equation}
	
\end{widetext}

\bibliography{ATRef}{}

\end{document}